\pdfoutput=1
\documentclass[twocolumn,10pt]{article}
\usepackage[margin=0.5in]{geometry} 
\usepackage{amsmath}
\usepackage{amsfonts}
\usepackage[mathscr]{euscript}
\usepackage{dsfont}
\usepackage{amssymb}
\usepackage{graphicx}
\usepackage{url}
\usepackage{rotating}
\usepackage{verbatim}
\usepackage{color}
\usepackage{svg}
\usepackage{phonetic}
\usepackage{cite}
\usepackage{hyperref}
\usepackage[font=footnotesize,labelfont=bf]{caption}
\usepackage{bm}
\usepackage{upgreek}

\usepackage[runin]{abstract}
\DeclareFontFamily{OT1}{pzc}{}
\DeclareFontShape{OT1}{pzc}{m}{it}{<-> s * [1.10] pzcmi7t}{}
\DeclareMathAlphabet{\mathpzc}{OT1}{pzc}{m}{it}

\newcommand{\ii}{\bm{\mathsf{i}}}
\newcommand{\interior}{\mathbf{i}}
\newcommand{\exterior}{\mathbf{c}}
\newcommand{\powerset}{\mathcal{P}}
\newcommand{\rset}{\mathcal{S}}

\newcommand{\motion}{\mathcal{T}}
\newcommand{\R}{\mathpzc{R}}
\newcommand{\F}{\mathcal{F}}

\newcommand{\volume}{V}
\newcommand{\area}{A}

\newcommand{\com}[1]{}

\newcommand{\blue}[1]{\textcolor{blue}{#1}}

\title{Haptic Assembly Using Skeletal Densities and Fourier Transforms%
    \footnote{This article was submitted on 07/11/2015 and published on 01/28/2016 in the ASME JCISE. For citation, please use:
    \protect\\
    \protect\\
        \blue{Behandish, Morad and Ilie\c{s}, Horea T., 2016. ``Haptic Assembly Using Skeletal Densities and Fourier Transforms.'' Journal of Computing and Information Science in Engineering, 16(2), p.021002.}
    \protect\\
    \protect\\
    A short version of this article was also presented in the ASME IDETC/CIE'2015 Conferences \cite{Behandish2015a}.
    }
}

\author{Morad Behandish and Horea T. Ilie\c{s}
\\
    {\small Departments of Mechanical Engineering and Computer Science and Engineering, University of Connecticut, USA}
}

\date{\small Technical Report No. CDL-TR-16-01, January 28, 2016}

\begin{document}

\maketitle

\noindent \hrule \vspace{5pt}
\begin{abstract}
Haptic-assisted virtual assembly and prototyping has seen significant attention over the past two decades. However, in spite of the appealing prospects, its adoption has been slower than expected. We identify the main roadblocks as the inherent geometric complexities faced when assembling objects of arbitrary shape, and the computation time limitation imposed by the notorious 1 kHz haptic refresh rate. We addressed the first problem in a recent work by introducing a generic energy model for geometric guidance and constraints between features of arbitrary shape. In the present work, we address the second challenge by leveraging Fourier transforms to compute the constraint forces and torques. Our new concept of `geometric energy' field is computed automatically from a cross-correlation of `skeletal densities' in the frequency domain, and serves as a generalization of the manually specified virtual fixtures or heuristically identified mating constraints proposed in the literature. The formulation of the energy field as a convolution enables efficient computation using fast Fourier transforms (FFT) on the graphics processing unit (GPU). We show that our method is effective for low-clearance assembly of objects of arbitrary geometric and syntactic complexity.
\end{abstract}

\vspace{5pt} \hrule \vspace{20pt}

\section{Introduction}

An integration of virtual reality (VR) tools into the modern computer-aided design (CAD) environments can facilitate an early-stage examination of a variety of product life-cycle aspects related to design, manufacturing, maintenance, service, and recycling. The so-called `virtual prototyping' \cite{Bullinger1999,Wang2002,Deviprasad2003} results in a significant reduction of time and cost associated with `physical prototyping', provides valuable insight into the functionality of the products, and facilitates the elimination of a large subset of design problems in the earlier stages of the process \cite{Bordegoni2006}. `Virtual assembly', defined as a simulated assembly of the virtual representations of mechanical parts in an immersive 3D user interface using natural human motions, characterizes an important subset of virtual prototyping \cite{Seth2006,Seth2008}, to which applying haptic feedback has been shown particularly beneficial in terms of task efficiency and user satisfaction \cite{Gomes1999,Volkov2001}.

In the past two decades, there have been numerous studies focused on virtual assembly using a variety of visualization tools (e.g., stereoscopic displays and goggles) and tracking devices (e.g., head tracking devices and data gloves) to assist the user in the assembly tasks. More recently, an increasing number of studies have leveraged haptic devices to provide a more realistic assembly experience with force feedback, a thorough survey of which is beyond the scope of this paper. We refer the reader to \cite{Seth2011} for an extensive review of previous studies, and to \cite{Vance2011,Perret2013} for recent insight on current knowledge and expected future directions in haptic assembly. Here we restrict ourselves to briefly overview the state-of-the-art practices and major recent contributions in Section \ref{sec_lit}, upon which we draw our newest approach for effective and efficient haptic assembly and disassembly.

Even after two decades replete with hundreds of important studies, the current models for real-time haptic-assisted assembly are still quite simplistic compared to the static and dynamic behavior of the objects in a real assembly process \cite{Perret2013,Behandish2014a,Behandish2015}. Even the most recent implementations are limited in number, complexity, and precision of assembly components. We aim to move on from few, simple, and high-clearance peg-in-hole examples to practical assembly scenarios with objects of complex geometric features and tight fits. This is hardly possible without revisiting the mathematical foundations for representing mechanical models and leveraging optimal algorithms to enable fast computations to comply with the 1 kHz requirement.

Among the major theoretical developments over the recent years in geometric computing are analytic methods \cite{Lysenko2010,Lysenko2013,Behandish2015d}. In contrast to their combinatorial counterparts that use tools from discrete geometry, analytic techniques take advantage of functions theory, convolution algebras, and harmonic analysis to solve fundamental problems related to detecting collisions, similarity, complementarity, and symmetry. The metrics for quantifying such notions are formulated as convolutions of shape descriptor functions. There are several important advantages associated with analytic modeling, namely:
\begin{itemize}
    \item It enables formulating and solving complex problems in a uniform fashion, whose treatments with classical methods are theoretically and computationally challenging. An important example is computing Minkowski operations \cite{Roerdink2000} on objects bounded by complex surfaces, which underlies swept volume generation \cite{Nelaturi2011} and collision detection \cite{Lysenko2013}, both of which are of prime importance in assembly planning (Section \ref{sec_cor}).
    \item It provides insight into systematic extension of the existing solutions to more difficult problems, e.g., formulating shape complementarity metrics \cite{Behandish2014} as a generalization of the collision predicate \cite{Lysenko2013} by using more sophisticated shape descriptors (Section \ref{sec_des}).
    \item It makes it possible to speed up the real-time computations to sub-millisecond times (required for haptic applications with 1 kHz frame rate) by using truncated Fourier transforms (Section \ref{sec_Fconv}).
    \item By working in the Fourier domain, the computational resources (i.e., time and memory) that are allocated to resolving part behavior in each frame can be altered on-the-fly by low-pass filtering, regardless of syntactic complexity (e.g., polygon count).
\end{itemize}
We aim to extend the application of analytic methods---whose use has been limited for the most part to robotics applications \cite{Lozano-Perez1983,Kavraki1995}---to both collision detection and geometric guidance for haptic assembly.

\subsection{Related Work} \label{sec_lit}

Over the past few years, a `two-phase' approach to haptic assembly has become popular in several different forms. The process is divided into a `free motion' phase in the presence of collision-induced physical constraints, and a `fine insertion' phase using pre-specified or computer-predicted geometric constraints \cite{Vance2011}.
Often the former is referred to as physically-based modeling (PBM), while the latter is called constraint-based modeling (CBM). The dynamic `part behavior' (mostly for PBM) is simulated by integrating the equations of motion in real-time using physics simulation engines (PSE) \cite{Chan2009,Sagardia2014} which provide a variety of functionalities (including collision response, impact/friction mechanics, and noon-smooth Lagrangian dynamics \cite{Renouf2005,Tching2008}).
The free motion phase typically relies on collision detection (CD) (reviewed in \cite{Lin1998,Jimenez2001,Kockara2007}) to constrain the motion with the so-called `physical constraints' (i.e., unilateral holonomic constraints) that automatically arise from the collision response, and to compute the necessary impact/contact forces to enforce those constraints \cite{Hasegawa2003,Hasegawa2004}, along with friction models \cite{Lotstedt1984,Stewart2000}. The fine insertion phase, on the other hand, is handled via constraint management libraries \cite{Marcelino2003,Murray2004} to implement additional `geometric constraints' (i.e., bilateral degree of freedom (DOF)-limiting constraints) that are manually prescribed or heuristically predicted based on simple geometric semantics \cite{Iacob2008,Iacob2011,Boussuge2012}.

Among the combinatorial CD methods that are popular in haptic assembly---due to their computational efficiency in the presence of the 1 kHz requirement---are Voronoi-clipping/marching methods (e.g., V-Clip \cite{Mirtich1998}, SWIFT \cite{Ehmann2000}, and SWIFT++ \cite{Ehmann2001}), oriented bounding box (OBB) tree-based methods (e.g, H-COLLIDE \cite{Gregory1999,Gregory2005} and RAPID \cite{Gottschalk1996}), and volumetric enumeration methods (e.g., the well-known Voxmap PointShell (VPS) \cite{McNeely2005,McNeely2006} used in the earlier versions of SHARP \cite{Seth2006,Seth2008}, and its improved variants \cite{Barbic2007,Sagardia2008}).
More efficient hierarchical data structures have been developed using spherical primitives (e.g., hierarchical bounding sphere (HBS) tree-based methods \cite{Hubbard1996,Bradshaw2004} and inner sphere tree (IST) methods \cite{Weller2009,Weller2011} successfully applied to haptic rendering \cite{Ruffaldi2008,Weller2009a,Weller2009b}). By leveraging the spherical symmetry in primitive collision predicates, sphere trees are generally more efficient alternatives to OBB trees, and represent nonuniform extensions to the uniform volumetric enumeration in VPS at every layer of the hierarchy. Although they have been shown to outperform VPS for nonconvex moving objects, their effectiveness to handle thin objects is yet to be tested \cite{Perret2013}.

In a separate line of research, analytic CD methods have been in use for decades in the robotics field for path planning in the presence of obstacles \cite{Lozano-Perez1983,Kavraki1995}. Unlike combinatorial methods that search for a collision certificate point (or lack thereof) in the intersection of the objects and recover a gradient-like quantity to evaluate the collision response, analytic methods formulate the collision constraint as a convolution of the objects' defining functions whose differentiation gives the collision response \cite{Lysenko2013}, both of which convert to simple algebraic operations in the Fourier domain. However, most FFT-based convolution methods have been presented for cumulative computation of the collision predicate for all configurations, which is more than what one needs in real-time VR applications with a single pose at question during each frame \cite{Lysenko2013}. In a recent work \cite{Lysenko2013}---which has been significantly influential on our development in Section \ref{sec_form}---the analytic CD was adapted to real-time applications along with explicit equations presented for gradient computations. Later, we proposed a hybrid combinatorial/analytic method in \cite{Behandish2015d} that used spherical decompositions similar to \cite{Hubbard1996,Bradshaw2004,Weller2009,Weller2011} to descretize the analytic equations, and demonstrated that sub-millisecond running times can be achieved with analytic CD regardless of input shape complexity. In spite of their great promise, the nascent real-time analytic methods are yet to be tested in VR applications, and to the best of our knowledge has not been implemented into PSEs.

Regardless of which CD method is used, PSE+CD alone has been found inadequate for low-clearance haptic assembly \cite{Seth2006,Seth2008}, partly due to the approximate nature of most CD methods (alleviated if exact boundary representations (B-reps) are used at the expense of computational performance, e.g., via the collision detection manager (CDM) module of D-Cubed in SHARP \cite{Seth2007,Seth2010}) and partly due to the noise in input data resulting from authentic hand vibration or added device errors \cite{Seth2011}.
This naturally led to the introduction of ad hoc solutions in a separate precision assembly phase, where a simplified set of `mating constraints' are explicitly introduced in close proximity of the final intended assembly configuration. The assembly constraints can be extracted from the CAD model (e.g., from ProE models in VADE \cite{Jayaram1999} and MIVAS \cite{Wan2004}) or specified on-the-fly (e.g., via the dimensional constraint manager (DCM) module of D-Cubed in SHARP \cite{Seth2007,Seth2010}). Rather than using the geometric semantics of the original parts, the virtual constraint guidance (VCG) method given in \cite{Tching2010a} relied on {\it manually} specified abstract geometric constructs referred to as `virtual fixtures' \cite{Rosenberg1993}---e.g., a pair of perpendicular planes intersecting at the axis of a cylindrical hole, to constrain and guide two points selected along the axis of a cylindrical peg.
Alternatively, the automatic geometric constraints (AGC) method investigated in \cite{Seth2010} and similar but independent studies in \cite{Iacob2008,Iacob2011,Boussuge2012} attempted to {\it automatically} identify the assembly intent and associated geometric constraints by matching `functional surfaces' \cite{Iacob2011}---e.g, a cylindrical surface characterized by its axis and diameter, which could be used to predict the intended mating relation and associated trajectories when a peg is brought to the proximity of a hole.

Up until recently, the two-phase framework has been widely accepted as the only promising future direction for performing haptic assembly tasks effectively \cite{Vance2011}. Among the main difficulties with this hybrid paradigm are detecting the proximity to an insertion site, properly switching between the two phases (i.e., `blending' algorithms to turn CD off and activate constraint management \cite{Seth2010}), and handling collision events outside the insertion site when CD is switched off for insertion \cite{Perret2013}. Furthermore, the ad hoc constraint management solutions are heavily dependent on user input based on {a priori} knowledge of the type of contact surfaces, which are in turn limited to simple categorized classes of geometric features---e.g., coplanarity of prismatic mates, coaxiallity of cylindrical joints, and alike.
In a similar classification to the one stated earlier for CD methods, the existing constraint identification algorithms can be identified as combinatorial methods that scale in running time with the complexity of the input CAD models, and look for certain heuristic indicators of a match between complementary features.

We recently introduced an alternative approach in \cite{Behandish2015} (a short version of which appeared in \cite{Behandish2014a}), which could be viewed as a unified analytic approach to computing predicates for collision response and geometric guidance. We defined a generic metric as a convolution that penalizes collisions and separations between objects, while it rewards shape complementarity of their mating features, resulting in an artificial geometric energy function that can be differentiated for force and torque computations.
As with most analytic approaches, our method generalizes to arbitrary geometry, without any simplifying assumption on the type or complexity of the mating features. Furthermore, it completely avoids the difficulties associated with the two-phase approach, along with the need for switching/blending between two distinct modes.
We demonstrated the effectiveness of the method for simple peg-in-hole benchmark examples, and tested haptic feedback in a 3 DOF setup. However, the mathematical complexity of the model in terms of volumetric integrals in the physical domain required time-consuming preprocessing steps as the geometric details of the parts and the number of DOF grew, introducing new challenges that we address in the present work by leveraging Fourier transforms.
%

\subsection{Contributions}

We propose a paradigm that streamlines assembly simulation using ideas from multivariate harmonic analysis \cite{Katznelson2004}.
The development expands upon our analytical formulation of the `geometric energy' field in \cite{Behandish2014a,Behandish2015}, defined as a convolution of the skeletal density functions (SDF) \cite{Behandish2014}.
The method subsumes analytic CD, and provides a generalization to analytic feature matching for geometric guidance.
The SDF shape descriptors are piecewise continuous functions defined over the 3D space for each individual part, whose distributions capture the topological and geometric properties of the surface features that partake in assembly.
In the context of assembly guidance, the SDF can be viewed as an implicit generalization of virtual fixtures mentioned earlier, whose computation is automated for arbitrary geometry.

The convolution in the physical space (where the part geometries reside) transfers into a pointwise multiplication of the Fourier expansion of the SDFs for the individual parts (i.e., the `amplitudes' of the multi-dimensional SDF signals). Guided by this property, we show that our previous formulation leads to a straightforward mathematical relationship between the Fourier representations of the SDF shape descriptors and the geometric energy field, which can benefit from the efficiency of the FFT algorithms \cite{Cooley1965}. Moreover, we present explicit analytic equations for computing the gradients of the convolution function (i.e., guidance forces and torques) for arbitrary spatial translations and rotations.
We implement the process using optimized FFT implementation on the highly-parallel GPU architecture. We show that haptic-enabled simulation of realistic assembly scenarios with complex CAD models and low-clearance fits is made possible to an adequate fidelity with the application of GPU-accelerated FFT calls.

\section{Formulation} \label{sec_form}

To compute the motion trajectory of objects in a virtual assembly scene, we aim to formulate an energy field over some configuration space (often abbreviated as the C-space) of the constituent parts, that
\begin{enumerate}
    \item penalizes collisions between the interiors of the pairs of objects, which is equivalent to an implicit representation of the so-called `configuration space obstacles' \cite{Lozano-Perez1983} (i.e., `ridges' of the energy terrain); and
    \item rewards proper fit/contact (i.e., configurations that exhibit superior shape complementarity between assembly features), which is equivalent to a re-scoring of the feasible collision-free C-space (i.e., `valleys' of the energy terrain).
\end{enumerate}
A recently proposed analytic approach to CD \cite{Lysenko2013} holds great promise for computing the narrowphase collision response in real-time for haptic assembly. The method defines a `gap function' between the two objects as a convolution of some implicit representation of the shapes---e.g., the `indicator function' (whose value is 1 for interior (and boundary) points, and 0 elsewhere) \cite{Kavraki1995}, distance-based depth functions, or smooth `bump functions' \cite{Lysenko2013}.
However, even {\it exact} CD is not sufficient for effective low-clearance insertion, since it computes large penalties for minor collision events that can occur due to hand vibration or device encoder inaccuracies \cite{Seth2010,Seth2011}. Furthermore, CD alone does not accomplish the second objective stated above, as will be detailed soon. By making use of a special class of shape descriptors (i.e., the SDF \cite{Behandish2014}) that better capture the geometric information of the assembly features, we overcome both problems in a unified framework.

\subsection{Preliminaries} \label{sec_prem}

Let us begin with a brief overview of the derivation of SDF descriptors elaborated in \cite{Behandish2014} and implemented in \cite{Behandish2015}. The different steps in the process can be identified as 1) a projection from the shape domain in the $3-$space to the complex domain $\mathds{C} \cong \mathds{R}^2$ composed of a Cartesian product of Euclidean metrics; 2) the application of a complex kernel to the projected boundary that extracts the topological and geometric properties pertinent to assembly; and 3) a surface integral over the boundary manifold to compute the shape descriptor. Here we give a different presentation that seems backwards in those steps, but provides useful insight into the subsequent Fourier analysis.\footnote{Although we encourage the reader to review the concepts in \cite{Behandish2015} for a more thorough perspective, it is not necessary for a comprehension of this paper, where the attempt has been to present the key concepts in a self-sufficient manner.}

Following the common conventions in solid modeling, we restrict our attention to the class of `well-behaved' solid objects $\rset \subset \powerset(\mathds{R}^3)$,\footnote{The collection $\powerset(A) = \{ B ~|~ B \subset A \}$ denotes the `power set' of a set $A$, i.e., the set of all subsets of $A$.}
defined as compact regular semi-analytic subsets of the Euclidean metric space with the usual topology based on the $L^2-$metric. This definition is general enough to encompass all practically significant shapes, yet sufficiently specific to avoid undesirable pathological behavior at the boundary \cite{Requicha1977a} and the skeleton \cite{Chazal2004}. Hereon, we use the common terms `solid' and `r-set' interchangeably to refer to a member of this class.

\begin{figure*}
    \centering
    \includegraphics[width=\textwidth]{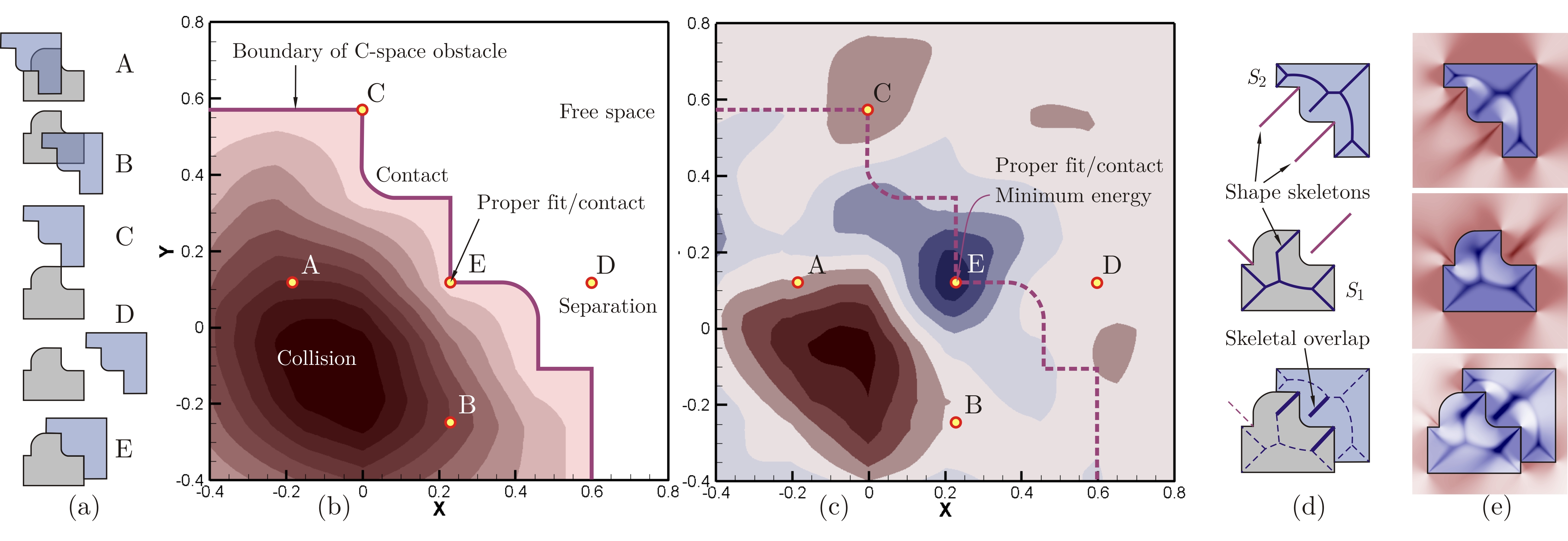}
    \caption{Different configurations (a) are evaluated using a simple gap function (b), and the shape complementarity score function (c), based on overlapping shape skeletons (d), formulated as a cross-correlation of skeletal densities (e).} \label{figure1}
\end{figure*}
%

\subsection{Correlation Paradigm} \label{sec_cor}

For a pair of arbitrary solids $S_1, S_2 \in \rset$, the configuration space of their relative rigid motions can be characterized with the special Euclidean group $\mathrm{SE}(3) \cong \mathrm{SO}(3) \ltimes \mathrm{T}(3)$ composed of a semi-direct product of the groups of proper orthogonal rotations $\R \in \mathrm{SO}(3)$ (represented by $3 \times 3$ orthogonal matrices $[\R]_{3 \times 3}$ with $\det(\R) = +1$), and translations $\mathbf{t} \in \mathrm{T}(3) \cong \mathds{R}^3$ (represented by arbitrary $3-$vectors $[\mathbf{t}]_{3 \times 1}$), together representing all possible rigid body transformations. In a virtual assembly environment, if $S_1$ and $S_2$ represent initial instances of the rigid solid parts (e.g., at rest on the assembly table), any subsequent position and orientation of the two parts (hereon referred to as the `absolute configurations') are given by $\motion_1 S_1$ and $\motion_2 S_2$, respectively, where $\motion_1, \motion_2 \in \mathrm{SE}(3)$ are arbitrary rigid motions (represented by tuples $\motion := (\R, \mathbf{t})$, or equivalently, $4 \times 4$ homogeneous matrices $[\motion]_{4 \times 4}$). The `relative configuration' of $S_2$ as observed from a coordinate frame fixed on $S_1$ is $\motion = \motion_1^{-1} \motion_2 \in \mathrm{SE}(3)$. Every such configuration can be conceptualized as a point in the 6D C-space $\mathrm{SE}(3)$. Following the common terminology in robotics and motion planning, the 6D geometric constructs that characterize the subsets of $\mathrm{SE}(3)$ corresponding to collisions between $S_1$ and $\motion S_2$ (or equivalently between $\motion_1 S_1$ and $\motion_2 S_2$ in the actual assembly scene) are called C-space obstacles (or `C-obstacles' for short), and their regularized complement in $\mathrm{SE}(3)$ is called the `free space'.

\paragraph{Cross-Correlation.}
Let $\rho_1, \rho_2 : \mathds{R}^3 \rightarrow \mathds{R}$ be real-valued functions over the $3-$space that implicitly define $S_1$ and $S_2$, respectively, i.e., $S_{1,2}$ are the regularized sublevel sets corresponding to $\rho_{1,2}(\mathbf{p}) \geq 0$ (or alternatively $\rho_{1,2}(\mathbf{p}) \leq 0$). Many important C-space problems can be formulated and solved in terms of the cross-correlation function $f_\mathrm{CC}: \mathrm{SE}(3) \rightarrow \mathds{R}$ defined as the following volume integral:
\begin{equation}
    f_\mathrm{CC}\big( \motion; S_1, S_2 \big) = \int_{\mathds{R}^3} \rho \big(\mathbf{p}; S_1 \big) \rho \big(\motion^{-1} \mathbf{p}; S_2 \big) d\volume, \label{eq_1}
\end{equation}
in which $\rho_{1,2}(\mathbf{p}) = \rho(\mathbf{p}; S_{1,2})$, where $\rho(\mathbf{p}; S)$ denotes a generic `defining function' for an arbitrary r-set $S \in \rset$; the integration variable is $\mathbf{p} = (x_1, x_2, x_3) \in \mathds{R}^3$ and $d\volume = dx_1 dx_2 dx_3$ is the volume element. Note that the defining function must be invariant under rigid body transformation, hence $\rho(\mathbf{p}; \motion S) = \rho(\motion^{-1}\mathbf{p}; S)$ for all $S \in \rset$ and $\motion \in \mathrm{SE}(3)$. Therefore, the integral in (\ref{eq_1}) overlaps the $\rho-$function of the two r-sets in their transformed positions and orientations. In order for $f_\mathrm{CC}( \motion; S_1, S_2)$ to remain bounded, the integrand functions need to either be compactly supported (e.g., $\rho(\mathbf{p}; S) = 0$ if $\mathbf{p} \notin S$) or approach zero as $\mathbf{p} \rightarrow \infty$ with a sufficiently rapid rate for the integral in (\ref{eq_1}) to converge.

\paragraph{Collision Detection.}
This correlation paradigm is central to important applications in geometric modeling and group morphology \cite{Roerdink2000,Lysenko2010}. For example, if we use the indicator function $\rho(\mathbf{p}; S) = 1$ if $\mathbf{p} \in S$ and $\rho(\mathbf{p}; S) = 0$ elsewhere, the integral in (\ref{eq_1}) simply computes the volume of the intersection $(S_1 \cap \motion S_2)$ (or equivalently, the volume of the intersection $(\motion_1 S_1 \cap \motion_2 S_2)$ of the moved pair of parts). Consequently, $f_\mathrm{CC} = 0$ characterizes the collection of
feasible configurations corresponding to zero intersection volume, including unassembled (no-contact) and assembled (proper point/curve/surface contact),
while $f_\mathrm{CC} > 0$ implicitly defines the C-space obstacles (i.e., regions of the C-space that correspond to an interpenetration of parts). The holonomic unilateral contact constraints can be implemented by penalizing the moving part's energy function proportional to the collision volume given by the so-called gap function $f_\mathrm{CC}(\motion; S_1, S_2)$, and the collision impulse forces and torques can be obtained from differentiating (\ref{eq_1}) using Lie algebras \cite{Lysenko2013}, to which we will return at the end of this section. Figure \ref{figure1} (b) illustrates the translational C-space landscape for a pair of 2D solids shown in panel (a), along with the colormap for the gap function. To make the illustration possible, the motion is restricted to translation only, i.e., the landscape in panel (b) is a section (corresponding to zero rotation) through the full 3D C-space. Each of the relative positions in panel (a) are represented by a point in panel (b). The gap function penalizes collision as in positions A and B ($f_\mathrm{CC} > 0$), but does not differentiate point contact in C and separation in D from proper fit/contact in E ($f_\mathrm{CC} = 0$). This is clearly due to the property $\rho(\mathbf{p}; S_{1,2}) = 0$ for $\mathbf{p} \notin S$ as a result of the definition.

\paragraph{Geometric Guidance.}
The question remains as how to modify this approach to 1) incorporate nonzero values to the cross-correlation function over the free space, such that it rewards proper fit/contact (e.g., position E in Fig. \ref{figure1} (a)), slightly penalizes separation and insufficient contact (e.g., positions C and D), in addition to the high penalty already assigned to collision (e.g., positions A and B); and 2) provide a mechanism to adjust relaxation of the collision penalty, and control the smoothness of the transition between the free space and the obstacle space. In other words, how can we add `valleys' to the free space in addition to the `ridges' inside the obstacle space, and how can we control the steepness of the transition between the two? The former allows additional haptic assistance for insertion, e.g., `magnetic' attraction forces and torques that guide the assembly by restricting the DOF of motion, and eventually snap the peg into the hole. The latter is important to let the user tune the flexibility of the collision response (i.e., `soft' versus `hard' assembly) as well as the degree of strictness with which the guidance constraints are enforced.

\begin{figure*}
    \centering
    \includegraphics[width=\textwidth]{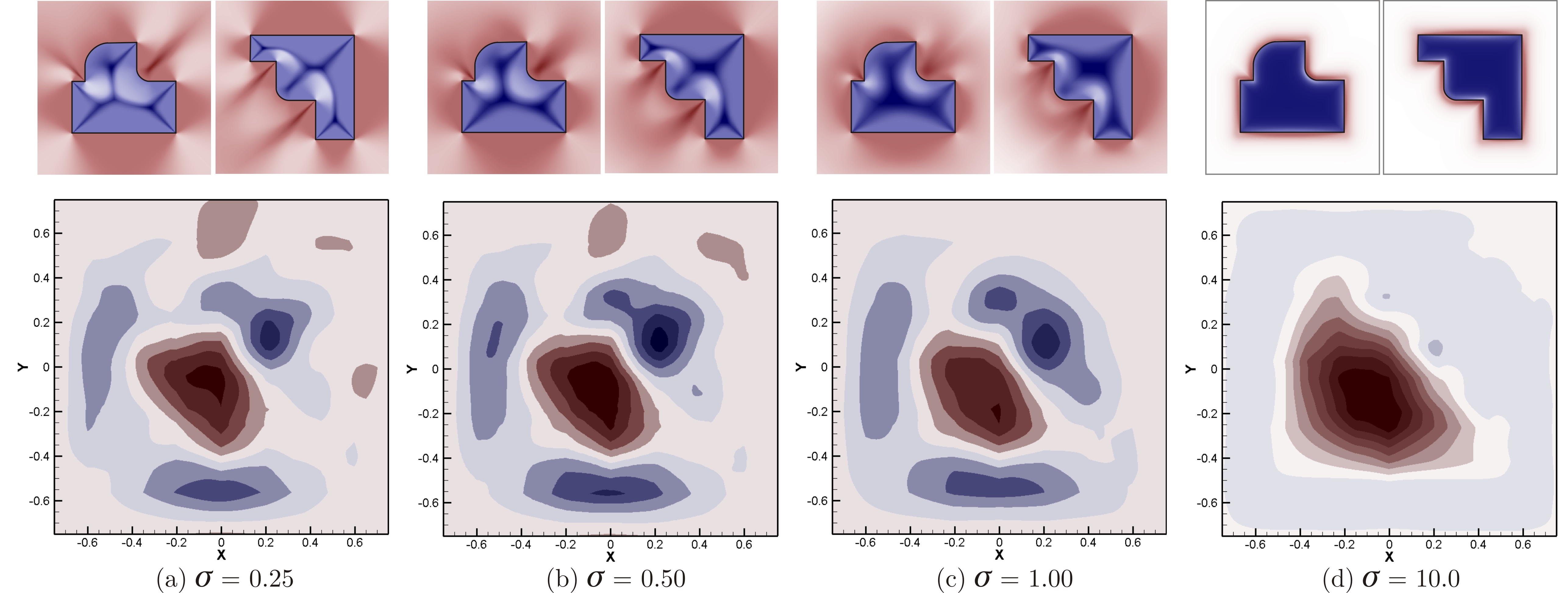}
    \caption{The extent of geometric details captured by the skeletal density distribution is adjustable by the thickness factor $\sigma$.} \label{figure2}
\end{figure*}
%

\subsection{Shape Descriptors} \label{sec_des}

The answers to both questions lie in the development of more sophisticated shape descriptors $\rho_{1,2} = \rho(\mathbf{p}; S_{1,2})$ with $\rho(\mathbf{p}; S_{1,2}) \neq 0$ outside the object, rather than using, for example, the binary indicator functions or bump functions.
We consider a special class of complex-valued functions $\rho_\sigma: (\mathds{R}^3 - \partial S) \rightarrow \mathds{C}$ that can be defined in the form of the following surface flux integral over the solid boundary \cite{Behandish2014,Behandish2015}:
\begin{equation}
    \rho_\sigma(\mathbf{p}; S) = \oint_{\partial S} \phi_\sigma \Big[ \zeta\left(\mathbf{p}, \mathbf{q}; S\right) \Big] ~ d\area_{\bot}, \label{eq_2}
\end{equation}
where $d \area_{\bot}$ is the signed area element normal to the line that connects $\mathbf{p} \in \mathds{R}^3$ to $\mathbf{q} \in \partial S$.
As detailed out in \cite{Behandish2015}, $\zeta: (\mathds{R}^3 \times \partial S) \rightarrow \mathds{C}$ characterizes a projection of the boundary $\partial S$ from the $3-$space to the complex plane, defined as
\begin{equation}
    \zeta(\mathbf{p}, \mathbf{q}; S) = \xi(\mathbf{p}; S) + \ii \eta(\mathbf{p}, \mathbf{q}),
\end{equation}
where the real-part $\xi(\mathbf{p}, S) = \pm \min_{\mathbf{q} \in \partial S} \| \mathbf{p} - \mathbf{q} \|_2$ is the signed Euclidean distance from the boundary $\partial S$ to the query point $\mathbf{p} \in \mathds{R}^3$,\footnote{The sign is positive if the query point is external ($\mathbf{p} \in \exterior S$) and negative if the query point is internal ($\mathbf{p} \in \interior S$), determined by Point Membership Classification (PMC) \cite{Klein2009}.}
and the imaginary-part $\eta(\mathbf{p}, \mathbf{q}) = \| \mathbf{p} - \mathbf{q} \|_2$ is simply the $L^2-$distance between one particular boundary point $\mathbf{q}$ to the point $\mathbf{p}$. The kernel $\phi_\sigma : (\mathds{C} - \{ 0 \}) \rightarrow \mathds{C}$ is the key, whose choice characterizes the set of topological and geometric properties that are extracted from the distance distribution embedded in the $\zeta-$map. The formulation as a surface integral characterizes a linear combination of the shares of different surface elements superimposed into $\rho_\sigma(\mathbf{p}; S)$.

It is interesting to note that the binary indicator functions used earlier to compute the collision gap function in (\ref{eq_1}) can be regarded as a special case of (\ref{eq_2}). If we let $\phi(\zeta) = -(4 \pi \eta^2)^{-1}$ where $\eta = \Im\{\zeta\}$ is the imaginary-part of the argument, the combination of the integrand and normal area element in (\ref{eq_2}) gives $\phi(\zeta) d\area_\bot = -d \gamma / 4\pi$ where $d \gamma = d\area_\bot / \| \mathbf{p} - \mathbf{q} \|_2^2$ is the signed infinitesimal spatial angle by which the query point $\mathbf{p} \in (\mathds{R}^3 - \partial S)$ observes the area element $d\area$ at $\mathbf{q} \in \partial S$. Therefore, the surface integral in (\ref{eq_2}) computes the `winding number,' which can be used for inclusion (i.e., PMC) testing \cite{Klein2009} (i.e., the winding number is 1 in the interior, 0 in the exterior, and undefined on the boundary of $S$). Disregarding the boundary points---which do not contribute to the volume integral in (\ref{eq_1})---this gives the indicator function whose convolution in (\ref{eq_1}) yields the gap function for the collision response.

\paragraph{Skeletal Density.}
We demonstrated in \cite{Behandish2014} that shape complementarity between objects of arbitrary shape can be related to the overlapping of shape skeleton branches that are generated by their assembly features---see Fig. \ref{figure1} (d), for example, where the assembly features (i.e., the two sharp corners and one filleted corner fitting together) are each captured by overlapping branches of the medial axis (MA). To develop an implicit continuous function that highlights the skeletal features, we first modify the aforementioned choice of the kernel as $\phi(\zeta) \propto \eta^{-2}$ (which led to the indicator function when substituted in (\ref{eq_2})), to $\phi_\sigma(\zeta) \propto \eta^{-2} g_\sigma (|\tan \angle \zeta| - 1)$, where $g_\sigma(x) = (\sqrt{2\pi} \sigma)^{-1} e^{-\frac{1}{2}(x/\sigma)^2}$ is the Gaussian function. This new term incorporates a higher contribution to the surface integral in (\ref{eq_2}) for boundary elements at which $\tan \angle \zeta \approx \pm 1$ (i.e., $\eta \approx |\xi|$) which means the boundary point $\mathbf{q} \in \partial S$ is an approximate nearest neighbor (ANN) to the query point $\mathbf{p} \in (\mathds{R}^3 - \partial S)$. Query points near the shape skeleton have more extensive ANNs, hence receive more such contributions. As depicted in Fig. \ref{figure1} (e), a continuous measure of skeletal overlap can be obtained by convolving such density functions using (\ref{eq_1}) for a given relative position and orientation of the two parts. The resulting change in the convolution function over the translational C-space is shown in Fig. \ref{figure1} (c), where the energy well is located at position E.
Figure \ref{figure2} (a--d) illustrates the effect of changing the thickness factor $\sigma > 0$ on the geometric energy landscape. Clearly, as $\sigma \rightarrow +\infty$ the Gaussian flattens out and the density function approaches the indicator function (with the exception of a signed coefficient), and as $\sigma \rightarrow 0^+$ the high density regions further resemble the MA.

For 3D interactions in haptic assembly, we used a slightly different complex structure $\phi(\zeta) \propto \zeta^{-2} g_\sigma (|\tan \angle \zeta| - 1)$ in \cite{Behandish2015} with different signed coefficients $\pm\lambda_{1,2}$ for the interior and exterior query points, determined by the PMC function. This allowed for positive real contributions to $f_\mathrm{CC}$ in (\ref{eq_1}) due to external-internal and internal-external skeletal overlaps (i.e., rewarding `proper fit'), and negative real contributions due to internal-internal and external-external overlaps (i.e., penalizing `collision' and `separation'), the relative intensities of which could be adjusted by the choice of the `penalty factor' $\mathfrak{p} = \lambda_2/\lambda_1$. %
We refer to the resulting generic function
$\rho(\mathbf{p}; S)$ as the `affinity function' in this context,
and to the cross-correlation function in (\ref{eq_1}) as the shape complementarity `score function' (denoted $f_\mathrm{SC}$ instead of $f_\mathrm{CC}$).
The geometric energy is then defined as $E_\mathrm{G} \propto \Re\{f_\mathrm{SC}\}$ and the guidance forces and torques are obtained as its gradients. Skipping the details, here it suffices to repeat that aligning the skeletal branches (as in Fig. \ref{figure1} (d)) can be thought of as a generalization of virtual fixtures \cite{Rosenberg1993} to arbitrary geometry, and implicit representation using SDF offers robustness and stability with respect to boundary perturbations, in contrast to the inherently unstable MA.

\subsection{Fourier Convolution} \label{sec_Fconv}

As a result of the definition, the SDF defined in (\ref{eq_2}) is invariant under rigid body motion. This means that for a rigid solid instance $S \in \rset$ and a transformation $\motion \in \mathrm{SE}(3)$, computing the $\rho-$function at $\mathbf{p} \in \mathds{R}^3$ for the moved instance $\motion S$ is equivalent to querying the precomputed SDF for the original instance at points displaced with the inverse transformation $\motion^{-1} \in \mathrm{SE}(3)$, i.e., $\rho_\sigma(\mathbf{p}; \motion S) = \rho_\sigma(\motion^{-1} \mathbf{p};  S)$.
Therefore, for a pair of solids $S_1, S_2 \in \rset$ transformed with $\motion_1, \motion_2 \in \mathrm{SE}(3)$, respectively, the score function per unit volume (whose real-part is referred to as the `geometric energy density') at a query point $\mathbf{p}' \in \mathds{R}^3 - (\partial(\motion_1 S) \cup \partial (\motion_2 S_2))$ can be computed from a product of the SDFs (measuring the skeletal overlap):

\begin{align}
    \frac{df_\mathrm{SC}}{d\volume}\Big|_{\mathbf{p}'} &= \rho_\sigma(\mathbf{p}'; \motion_1 S_1) ~\rho_\sigma(\mathbf{p}'; \motion_2 S_2) \\
    &= \rho_\sigma(\mathbf{p}; S_1) ~\rho_\sigma(\motion_2^{-1} \motion_1 \mathbf{p}; S_2).
\end{align}
where $\motion = \motion_1^{-1} \motion_2 \in \mathrm{SE}(3)$ is the motion of $S_2$ as observed from a coordinate frame attached to $S_1$ (compare with (\ref{eq_1})), and $\mathbf{p} = \motion_1^{-1} \mathbf{p}'$ is the new coordinates of the query point measured with respect to that frame. The overlap density $d f_\mathrm{SC}/d\volume$ is then accumulated over the $3-$space to obtain the score in (\ref{eq_1}).

\paragraph{Motion Decomposition.}
The only time-dependent variable in (\ref{eq_1}) is the 6D relative motion $\motion \in \mathrm{SE}(3)$.
To simplify the subsequent development, let us decompose the motion into the translational component $\mathbf{t} \in \mathrm{T}(3)$ described by a $3-$tuple $(\mathpzc{t}_1, \mathpzc{t}_2, \mathpzc{t}_3) \in \mathds{R}^3$, and the rotational component $\R \in \mathrm{SO}(3)$ represented by a $3\times3$ proper orthogonal matrix $[\R]_{3\times3}$---`orthogonal meaning $\R^{-1} = \R^\mathrm{T}$ and `proper' meaning $\mathrm{det}(\R) = +1$.
As a result of the product structure $\mathrm{SE}(3) = \mathrm{SO}(3) \ltimes \mathrm{T}(3)$, the transformation sequence applies as $\motion \mathbf{p} = (\R \mathbf{p}) + \mathbf{t}$ hence $\motion^{-1} \mathbf{p} = \R^\mathrm{T} (\mathbf{p} - \mathbf{t})$.
Noting the rotational invariance property of the SDF, it is easy to see that for all $S \in \rset$,
\begin{equation}
    \rho_\sigma(\motion^{-1} \mathbf{p}; S) = \rho_\sigma(\R^\mathrm{T} (\mathbf{p} - \mathbf{t}); S) = \rho_\sigma(\mathbf{p} - \mathbf{t}; \R S).
\end{equation}
Using the notation $\rho_{1,2}(\mathbf{p}) = \rho_\sigma(\mathbf{p}; S_{1,2})$ and defining the reflection of a generic function as $\tilde{f}(\mathbf{p}) = f(-\mathbf{p})$, the integral in (\ref{eq_1}) becomes a 3D convolution over a section through the C-space corresponding to the fixed rotation $\R \in \mathrm{SO}(3)$:
\begin{align}
    f_\mathrm{SC}((\R,\mathbf{t}); S_1,S_2) &= \int_{\mathds{R}^3} \rho_1(\mathbf{p}) \tilde{\rho}_2 \left(\R^\mathrm{T}(\mathbf{t} - \mathbf{p}) \right) d\volume \nonumber \\
    &= \left(\rho_1 \ast (\tilde{\rho}_2 \circ \R^\mathrm{T}) \right)(\mathbf{t}). \label{eq_3}
\end{align}

\paragraph{Fourier Transforms.}
Using the orthonormal Fourier basis of the form $e^{2\pi \ii(\bm{\upomega} \cdot \mathbf{p})}$ where $\mathbf{p}, \bm{\upomega} \in \mathds{R}^3$, one could decompose a function $f: \mathds{R}^3 \rightarrow \mathds{C}$ (defined over the 3D physical domain) into its components captured by the function $\hat{f}: \mathds{R}^3 \rightarrow \mathds{C}$ (defined over the 3D frequency domain). The forward Fourier transform $\hat{f} = \F\{ f \}$ is thus defined as the inner product
\begin{equation}
    \hat{f}(\bm{\upomega}) = \left\langle f, e^{+2\pi \ii(\bm{\upomega} \cdot \mathbf{p})} \right\rangle = \int_{\mathds{R}^3} f(\mathbf{p}) e^{-2\pi \ii(\bm{\upomega} \cdot \mathbf{p})} d\volume, \label{eq_4a}
\end{equation}
and the inverse Fourier transform $f = \F^{-1} \{ \hat{f} \}$ is defined as follows, to retrieve the function as a superposition of its orthogonal components:
\begin{equation}
    f(\mathbf{p}) = \left\langle \hat{f}, e^{-2\pi \ii(\bm{\upomega} \cdot \mathbf{p})} \right\rangle = \int_{\mathds{R}^3} \hat{f} (\bm{\upomega}) e^{+2\pi \ii(\bm{\upomega} \cdot \mathbf{p})} d\volume. \label{eq_4b}
\end{equation}
The integrals in (\ref{eq_4a}) and (\ref{eq_4b}) are sometimes referred to as the continuous Fourier transform (CFT), whose discredited form for computation purposes is given in Section \ref{sec_DFT}.
The {\it convolution theorem} states that the Fourier transform of a convolution is the pointwise product of Fourier transforms \cite{Katznelson2004}, hence the integral in (\ref{eq_3}) converts in the frequency domain to the simple product $\hat{f}_\mathrm{SC}(\bm{\upomega}) = \F \{ f_\mathrm{SC} \} = \F \{\rho_1\} \F \{(\tilde{\rho}_2 \circ \R^\mathrm{T})\}$. As a direct consequence of the definition in (\ref{eq_4a}), the rotation and reflection commute with the Fourier transform (noting that $\bm{\upomega} \cdot (\R^\mathrm{T} \mathbf{p}) = (\R \bm{\upomega}) \cdot \mathbf{p}$ and $\bm{\upomega} \cdot (-\mathbf{p}) = (-\bm{\upomega}) \cdot \mathbf{p}$ in (\ref{eq_4a})), hence $\F\{\tilde{\rho}_2 \circ \R^\mathrm{T}\} = \F\{\tilde{\rho}_2\} \circ \R^\mathrm{T}$ in which $\F \{\tilde{\rho}_2\} = \tilde{\F}\{\rho_2\}$. The score function in the physical domain can thus be computed by applying an inverse transform as
\begin{equation}
    f_\mathrm{SC}((\R,\mathbf{t}); S_1,S_2) = \F^{-1} \left\{ \F\{\rho_1\} \left(\F \{\tilde{\rho}_2\} \circ \R^\mathrm{T} \right) \right\}. \label{eq_5}
\end{equation}
At a first glance, this might appear as aggravating the computational burden, by requiring the evaluation of 3 volume integrals rather than the one in (\ref{eq_3}). However, (\ref{eq_4a}) and (\ref{eq_4b}) can be computed efficiently using the FFT algorithm \cite{Cooley1965}, as will be demonstrated in Section \ref{sec_imp}.

\paragraph{Convolution Gradients.}
The forces and torques can be obtained from the gradient of the geometric energy field $E_\mathrm{G}: \mathds{R}^3 \rightarrow \mathds{R}$, which is proportional to the real-part of the shape complementarity score function $f_\mathrm{SC}: \mathds{R}^3 \rightarrow \mathds{C}$, i.e., its rate of change with respect to infinitesimal translations and rotations. This is possible by differentiating either the physical domain formulation (\ref{eq_3}) (detailed in \cite{Behandish2015}) or the frequency domain formulation in (\ref{eq_5}).
Another important advantage of working in the Fourier domain is that the translational differentiation is replaced with a multiplier:
\begin{equation}
    \langle \frac{d f_\mathrm{SC}}{d \mathbf{t}}, \mathbf{e} \rangle = (2\pi \ii) \F^{-1} \Big\{ (\mathbf{e} \cdot \bm{\upomega}) \F\{\rho_1\} \left(\F \{\tilde{\rho}_2\} \circ \R^\mathrm{T} \right)  \Big\}, \label{eq_6a}
\end{equation}
where $\mathbf{e} \in \mathds{R}^3$ represents any direction in the vector space $\mathrm{T}(3) \cong \mathds{R}^3$, along which the differentiation occurs. The rotational differentiation, on the other hand, is more involved since $\mathrm{SO}(3)$ is not a vector space and cannot be globally parameterized by a single continuous 3D grid. To obtain a local parametrization, the tangent direction at $\R \in \mathrm{SO}(3)$ is obtained as $\R \Omega$ where $\Omega \in \mathfrak{so}(3)$ can be represented by a $3 \times 3$ skew-symmetric matrix $[\Omega]_{3\times3}$, and $\mathfrak{so}(3)$ denotes the Lie algebra for $\mathrm{SO}(3)$, which is a vector space tangent to $\mathrm{SO}(3)$ at the identity rotation \cite{Lysenko2013}. Without getting into much detail, we present the rotational gradient as
\begin{equation}
    \langle \frac{d f_\mathrm{SC}}{d \R}, \mathbf{e} \rangle = (2\pi \ii) \F^{-1} \Big\{ (\R^\mathrm{T} \Omega \bm{\upomega}) \cdot \F\{\rho_1\} \left( \F \{\tilde{\bm{\uprho}}^\ast_2 \} \circ \R^\mathrm{T} \right) \Big\}, \label{eq_6b}
\end{equation}
where $\mathbf{e} \in \mathds{R}^3$ is the dual vector of $\Omega \in \mathfrak{so}(3)$, which means $\Omega \bm{\upomega} = \mathbf{e} \times \bm{\upomega}$. $\bm{\uprho}^\ast_2 : \mathds{R}^3 \rightarrow \mathds{C}^3$ is a vector function defined as $\bm{\uprho}^\ast_2 (\mathbf{p}) = +\rho_2 (\mathbf{p}) \mathbf{p}$ (hence $\tilde{\bm{\uprho}}^\ast_2 (\mathbf{p}) = -\tilde{\rho}_2 (\mathbf{p}) \mathbf{p}$), whose Fourier transform $\F\{\tilde{\bm{\uprho}}^\ast_2\}$ can be obtained by componentwise FFTs of its complex components projected on the 3 coordinate axes. See \cite{Lysenko2013} for a more in-depth treatment of rotational gradients.

The 3D translational and rotational gradient vectors can be computed in a componentwise fashion by substituting for the base vectors $\mathbf{e} \in \{\mathbf{e}_1, \mathbf{e}_2, \mathbf{e}_3 \}$ one at a time in (\ref{eq_6a}) and (\ref{eq_6b}).
The complete 6D gradient $\nabla f_\mathrm{SC} : \mathrm{SE}(3) \rightarrow \mathds{C}^6$ is defined as $\nabla f_\mathrm{SC} = (df_\mathrm{SC} / d\mathbf{t}, df_\mathrm{SC} / d\R)$, and the geometric force and torque are obtained as $(\mathbf{F}_\mathrm{G}, \mathbf{T}_\mathrm{G}) = - \nabla E_\mathrm{G} \propto \Re \{ \nabla f_\mathrm{SC} \}$ as detailed in \cite{Behandish2015}.

\section{Implementation} \label{sec_imp}

In this section we present the numerical algorithms that carry out a discrete approximation of the SDF integrals and the subsequent Fourier transforms for a particularly simple representation (namely, triangular mesh B-reps). We also present a straightforward complexity analysis of each step, and refer the reader to \cite{Behandish2014} for more details.

\subsection{Precomputations} \label{sec_rep}

The following summarizes the offline SDF precomputations for mesh representations in both physical and frequency domains, the former being described in more detail in \cite{Behandish2015}.

\paragraph{Representation.}
The analytic definition of the SDF in (\ref{eq_2}) does not impose any restriction on the representation scheme, as long as it satisfies the informational completeness requirement \cite{Requicha1980a}, and more specifically, supports distance and inclusion queries for arbitrary query points in the 3D space.  Nevertheless, the numerical computation of the surface integral in (\ref{eq_2}) lends itself well to B-reps, and particularly to triangular mesh approximations. For mesh generation from STEP models (exported from CAD software of choice), we use the NETGEN library \cite{Schoberl1997}.
We obtain the unsigned distance field using Havoc3D \cite{Hoff1999}, which computes the rasterized Voronoi diagram using the OpenGL rendering pipeline and depth-buffer.
To compute the PMC and correct the sign of the distance function, on the other hand, we take advantage of the winding number approach in \cite{Klein2009}, which can be thought of as a special case of (\ref{eq_2})---i.e., one with a simple inverse-square kernel $\phi(\xi + \ii \eta) = (4\pi \eta^2)^{-1}$ as explained earlier in Section \ref{sec_des}---which can be computed using the same subroutines that compute the SDF with arbitrary kernels.

\paragraph{Preprocessing.}
The sequence of steps can be summarized as
1) generating a mesh from the CAD model using NETGEN \cite{Schoberl1997};
2) computing the unsigned distance function using Havoc3D \cite{Hoff1999};
3) evaluating the PMC function \cite{Klein2009} (to correct the distance signs) using (\ref{eq_2}) with an inverse-square kernel; and
4) evaluating the SDF descriptor \cite{Behandish2014} using (\ref{eq_2}) with a combined Gaussian and inverse-square kernel.
The last two steps can be implemented by approximating the integral in (\ref{eq_2}) as a discrete Riemann sum over the triangles. In order to ensure numerical stability one must guarantee an upperbound on the error, which is not possible by assigning a lumped weight to each triangle in the sum. This is due to the inverse-square term in the kernel, resulting in large errors as the query point gets closer to a particular triangle. To overcome this difficulty, the algorithm carries out adaptive recursive subdivisions of the triangles with a spatial angle-based threshold, i.e., until the triangle is subdivided to small enough pieces each observed from the query point by a small spatial angle $\delta \gamma = \delta A_\bot /(4\pi \eta^2)$.
More implementation details for SDF precomputation in the physical domain are given in \cite{Behandish2015}, and will not be repeated here. It suffices to mention that for a mesh $\Delta_n(S) = \bigcup_{j = 1}^n \delta_j$ that approximates the boundary $\partial S$ with the complex of $n$ faces $\delta_j~(1 \leq j \leq n)$ and $O(n)$ vertices and edges, and a uniformly sampled 3D grid of $m$ query points $G_m(S)$ that contains the bounding box of $S$, all steps take $O(mn)$ sequential time and $O(m+n)$ memory space.
Furthermore, the summation can be parallelized in a trivial manner by assigning disjoint subsets of the grid to different processors.

We implement (\ref{eq_2}) in parallel for the multi-core central processing units (CPU) using the Boost C++ libraries \cite{Schling2011} for multi-threading,\footnote{In the earlier implementation reported in \cite{Behandish2015} we used OpenMP for the task of CPU multi-threading. A significant improvement in performance was observed by reimplementing the same steps using the Boost C++ \cite{Schling2011} `thread' library.}
and obtain significant speed-ups that scale almost linearly with the number of dedicated cores. In addition, the 3D grid structure maps properly to the single-instruction multiple-thread (SIMT) execution model of the modern many-core GPUs, allowing us to further speed up the process using NVIDIA's compute-unified device architecture (CUDA).

\begin{figure*}
    \centering
    \includegraphics[width=\textwidth]{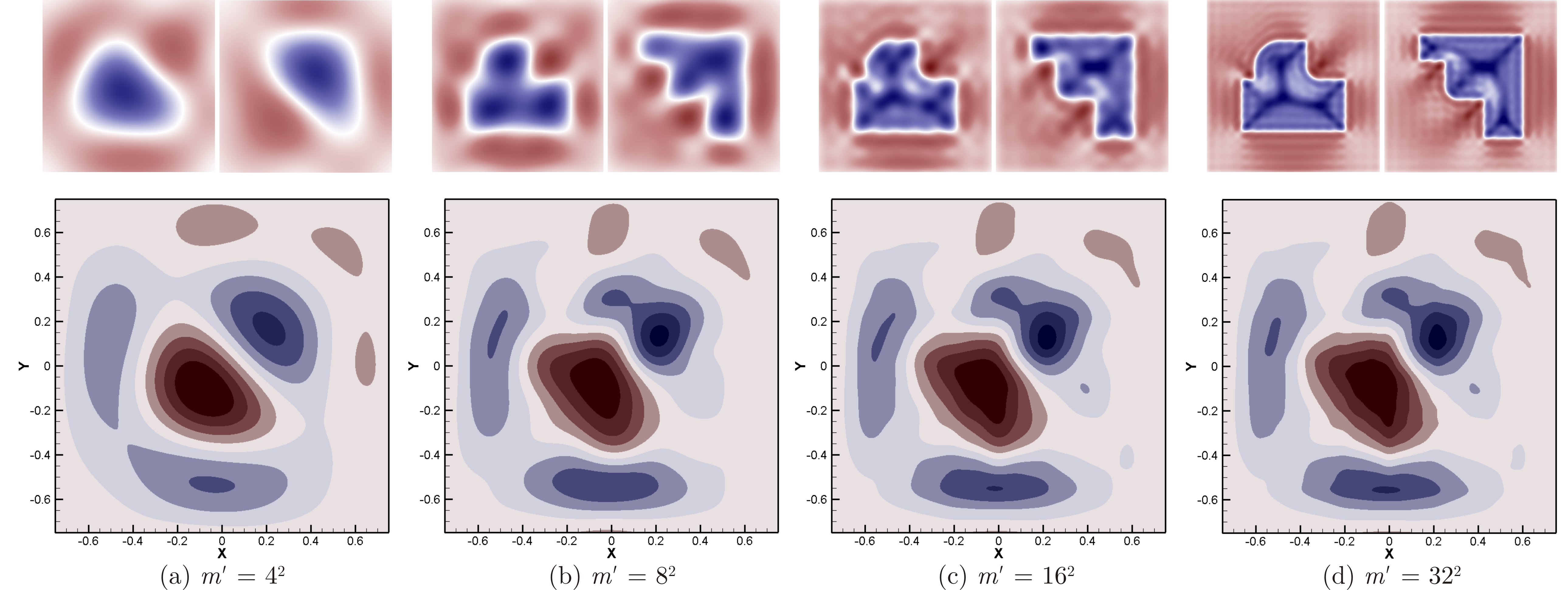}
    \caption{Frequency domain representation allows for a systematic means of successive approximation of the energy field.} \label{figure3}
\end{figure*}

\subsection{FFT Computations} \label{sec_DFT}

By sampling the physical and frequency domain data over the 3D grids $G_m(S)$ and $\hat{G}_m(S)$, respectively, the volume integrals in (\ref{eq_4a}) and (\ref{eq_4b}) that define the forward and inverse CFTs, respectively, can be approximated by the sums
\begin{align}
    \hat{f}_k &\approx \sum_{i = 1}^m f_i e^{-2\pi \ii(\bm{\upomega}_k \cdot \mathbf{p}_i)} \delta\volume, \quad 1 \leq k \leq m, \label{eq_7a}\\
    f_i &\approx \sum_{k = 1}^m \hat{f}_k e^{+2\pi \ii(\bm{\upomega}_k \cdot \mathbf{p}_i)} \delta\volume, \quad 1 \leq i \leq m, \label{eq_7b}
\end{align}
where $\mathbf{p}_i \in G_m(S)$ and $\bm{\upomega}_k \in \hat{G}_m(S)$ are uniformly sampled physical and frequency nodes at which the function values $f_i := f(\mathbf{p}_i)$ and $\hat{f}_k := \hat{f}(\bm{\upomega}_k)$ are stored, respectively. With the exception of a constant factor (depending on the chosen conventions), these sums define the discrete Fourier transform (DFT) whose cumulative computation for all grid nodes takes $O(m^2)$ basic operations using the cascade method. However, the same computation can be carried out in $O(m \log m)$ steps using the radix-2 FFT algorithm \cite{Cooley1965}.

A key observation is that the frequency domain representations of the SDFs, namely $\hat{\rho}_1 = \F \{\rho_1\}$, $\hat{\rho}_2 = \F \{\rho_2\}$, and $\hat{\bm{\uprho}}^\ast_2 = \F \{\bm{\uprho}^\ast_2\}$ depend on part geometries alone (and not on the instantaneous assembly configuration), hence can be precomputed offline prior to the virtual assembly session.
For a pair of parts, computing the forward FFT in (\ref{eq_7a}) to obtain $\hat{\rho}_{1,k}, \hat{\rho}_{1,k}, \hat{\bm{\uprho}}^\ast_{2,k}~(1 \leq k \leq m)$ from $\rho_{1,i}, \rho_{2,i}, \bm{\uprho}^\ast_{2,i}~(1 \leq i \leq m)$ takes $O(m \log m)$ per part.

The subsequent computation of the convolutions in (\ref{eq_5}), (\ref{eq_6a}), and (\ref{eq_6b}) (for geometric energy, force, and torque evalutations) in real-time for a particular relative orientation $\R \in \mathrm{SO}(3)$ of the parts takes place entirely in the frequency domain. The sequence of operations is 1) interpolating $(\hat{\rho}_2 \circ \R^\mathrm{T})$ and $(\hat{\bm{\uprho}}^\ast_2 \circ \R^\mathrm{T})$ over a rotated grid $\R^\mathrm{T} \hat{G}_m(S_2)$, followed by a reflection, which takes $O(m)$ basic trilinear interpolation steps; 2) a pointwise multiplication of the interpolated data with $\hat{\rho}_1$ data over $\hat{G}_m(S_1)$, which also takes $O(m)$ basic steps; and 3) an inverse FFT along with applying the proper coefficients (e.g., $2\pi \ii$ for force/torque computations), which takes $O(m \log m)$ steps and yields the convolution results over a grid of translations. Therefore, the total arithmetic complexity of the process (for a single rotation and all translations of interest) is $O(m \log m)$.

We use the FFTW library \cite{Frigo2005} for the CPU sequential implementation and NVIDIA's cuFFT(W) for the GPU parallel implementation of the FFT, the running times of both to be presented in Section \ref{sec_on} for comparison.

\subsection{Low-Pass Filtering} \label{sec_filt}

The additional important advantage of this representation is that one could decide to keep only a small subset of $m' \ll m$ frequency domain data nodes (i.e., the dominant modes) for computing the Fourier convolutions in (\ref{eq_5}), (\ref{eq_6a}), and (\ref{eq_6b}) in real-time, in a trade-off between the desired accuracy and available computational power. This results in a reduction of the real-time process complexity to $O(m' \log m')$ operations per frame, which is practically almost $O(1)$.
In other words, the interpolation (of the rotated and reflected data), the pointwise multiplication, and the inverse FFT steps are carried out for a significantly smaller sample of low-frequency grid nodes, a process referred to as `low-pass filtering'. As we will demonstrate in Section \ref{sec_on}, the pointwise multiplication step is the bottleneck and dictates an upperbound on the choice of $m'$ in order to stay within a prespecified time allocated to guidance force and torque computations at each frame, hence to achieve the desired frame rate of 1 kHz.

It is worthwhile noting that the inverse FFT cumulatively computes the correlations for a range of $m$ translations corresponding to a single rotation. This is a nonoptimal approach from a theoretical point of view, since we only need the results for a single configuration, i.e., the instantaneous relative translation and rotation of the objects $(\R, \mathbf{t}) \in \mathrm{SE}(3)$ at the current simulation frame. Therefore, $O(m' \log m')$ of the FFT can be reduced to $O(m')$ of a cascade partial summation of (\ref{eq_7b}) for a single translation index $i$. However, this does not yield a significant performance gain in practice, where the running time is governed by the notably slower pointwise multiplication step. Besides, parallel implementation of a cascade sum (especially on the GPU) for small $m'$ does not necessarily outperform its FFT counterpart.

Figure \ref{figure3} (a--d) illustrates the successive Fourier approximations for the 2D skeletal densities in Fig. \ref{figure2} (b) with $\sigma = 0.5$. A grid size of $m = 512^2 = $ 262,144 nodes is used, resulting in the same number of frequency domain amplitudes. In this case, the evolution of the score function is not substantial for $m' > 16^2 = 256$ (less than $0.1\%$ of $m$).
However, shapes with higher geometric detail require a larger number of modes to capture the smallest features.
It is very important to note that the input model complexity (e.g., the number of triangles $n_{1,2}$ per mesh representation of the solids $S_{1,2} \in \rset$) is irrelevant here, and the suitable value of $m'$ is determined from the desired fidelity with which the small geometric features are captured in the output energy field. This indifference to the syntactic representation complexity is a significant advantage of our method over most other collision detection and constraint management algorithms, whose running times depend on (and scale with) the representation size, resulting in a failure to handle large mesh sizes in real-time due to the high frame rate limitation.

\begin{figure}
    \centering
    \includegraphics[width=0.48\textwidth]{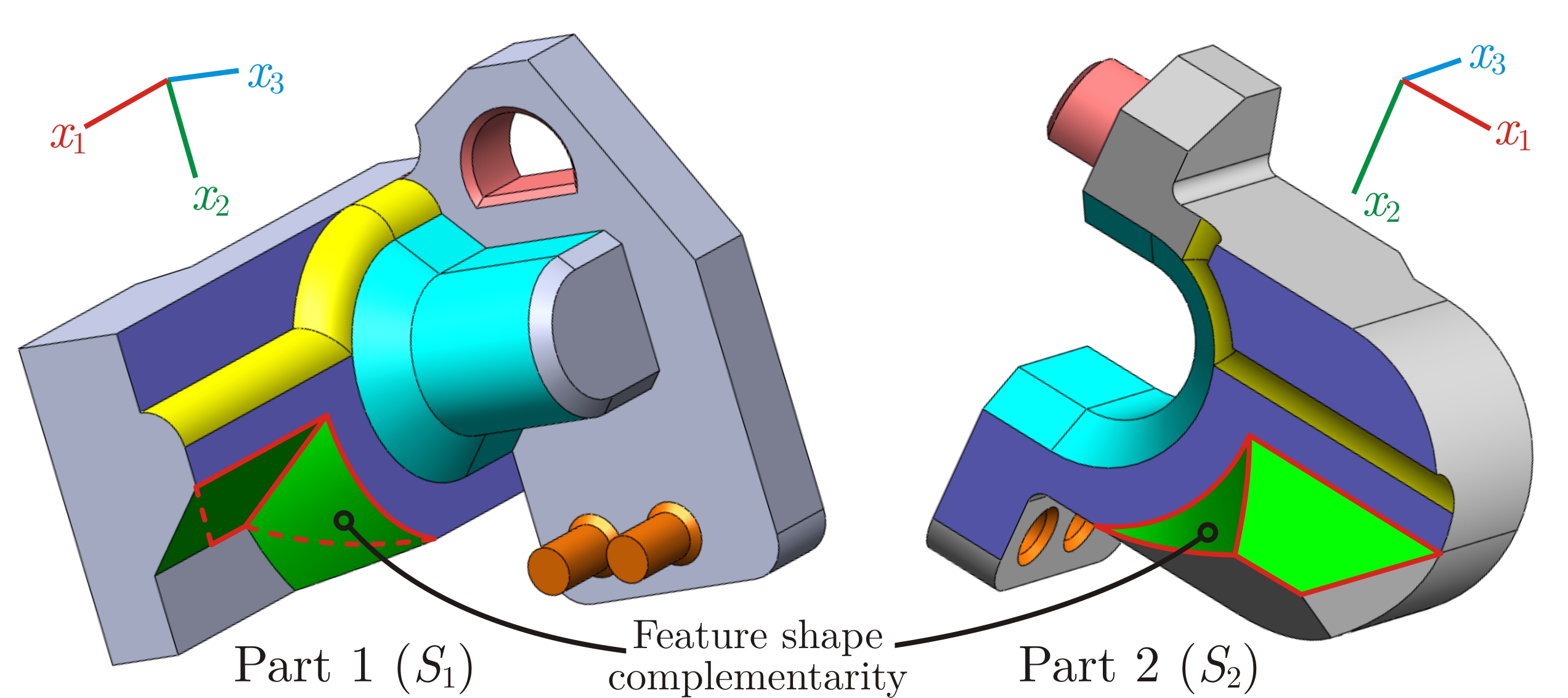}
    \caption{A non-trivial, zero-clearance assembly pair.} \label{figure4}
\end{figure}
%

\section{Results \& Discussion}

In \cite{Behandish2015}, we tested the SDF descriptors for haptic-assisted assembly of simple peg-in-hole geometries. The examples provided important insight into the properties of SDF and its effectiveness as an automated generalization of virtual fixtures for arbitrary geometry. However, the numerically intensive cascade computation of the integral in (\ref{eq_1}) required substantial preprocessing time. Here we use the FFT convolution technique, using optimized CPU- and GPU-accelerated implementations. To demonstrate the practicality of the method, we use the pair of 3D assembly parts shown in Fig. \ref{figure4}. The solids in this example are made of semi-algebraic r-sets with only planar, cylindrical, spherical, and toroidal surface patches, which forms a small subset of the general semi-algeberaic class (i.e., solids bounded by polynomial surfaces of arbitrary degrees) and even more general semi-analytic class covered by our formulation. Nevertheless, an automatic identification of the correspondence between the mating features (depicted with different colors) is not trivial from an algorithmic perspective---e.g., recognition and matching of the partially complementary features connected by a curve. Furthermore, there are 3 pairs of pegs and holes with zero clearance, making this example sufficiently challenging.

\begin{figure*}
    \centering
    \includegraphics[width=\textwidth]{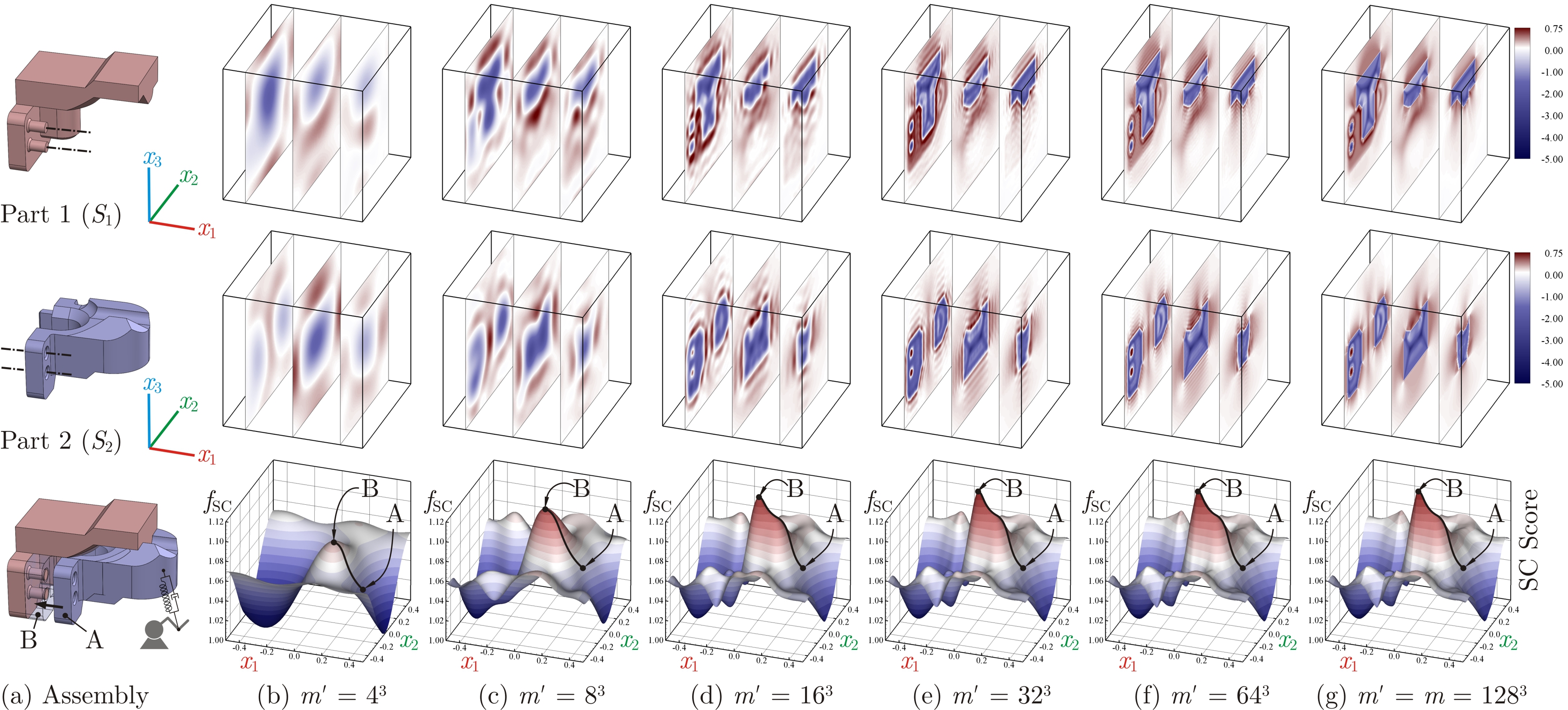}
    \caption{The effect of FFT filtering on part SDFs (top) and score variations versus biaxial relative translation (bottom).} \label{figure8}
\end{figure*}
%

\subsection{Offline Preprocessing} \label{sec_off}

\begin{figure}
    \centering
    \includegraphics[width=0.48\textwidth]{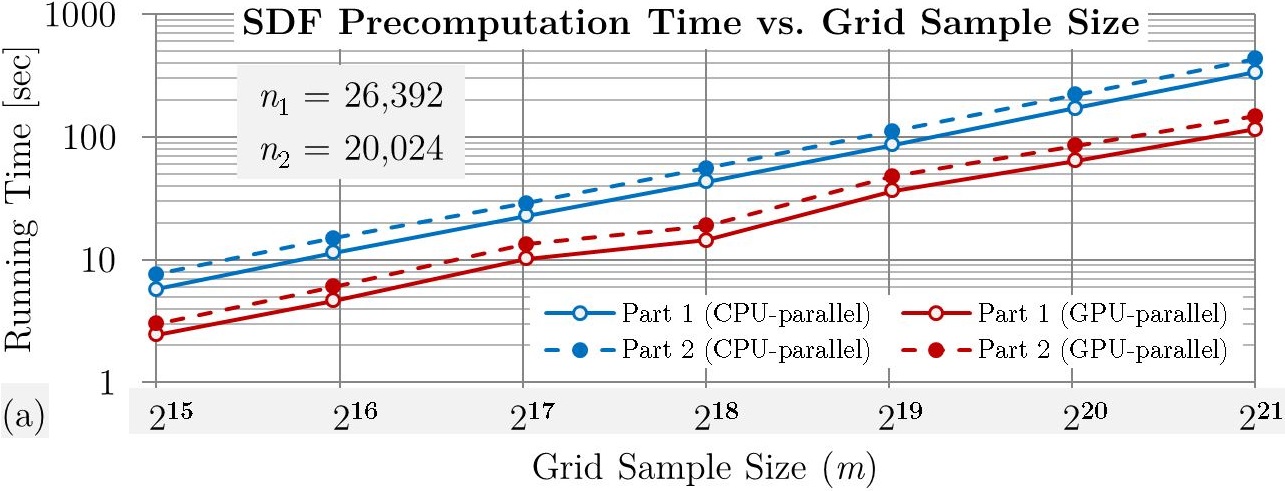}
    \includegraphics[width=0.48\textwidth]{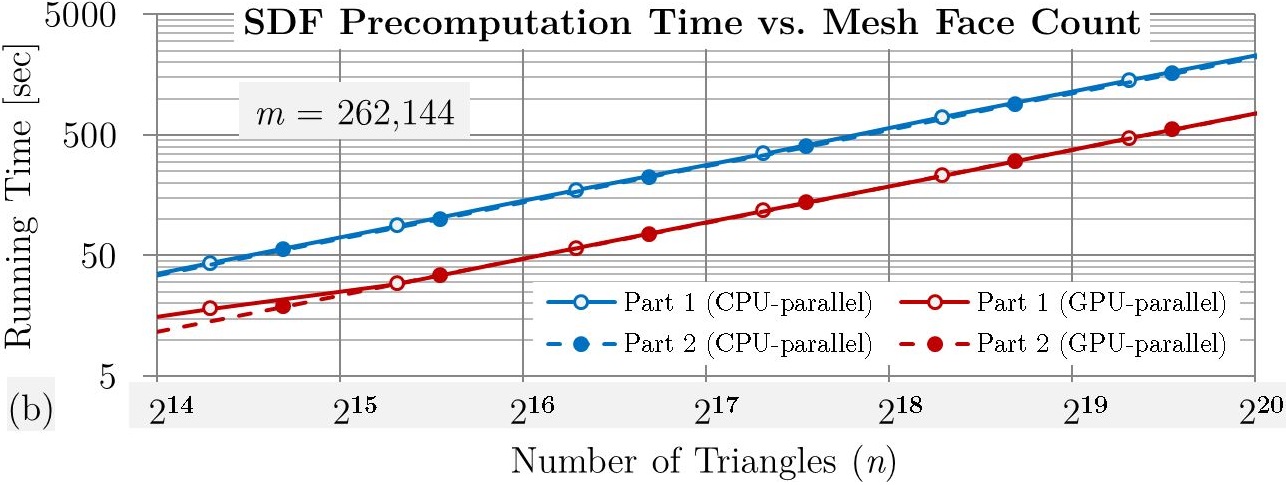}
    \caption{CPU vs. GPU performances for SDF computation.} \label{figure5}
\end{figure}

The running times for the offline steps made of precomputing the part SDFs in the physical domain and their forward FFT into the frequency domain are plotted in Figs. \ref{figure5} and \ref{figure6}.

\paragraph{Precomputing SDFs.}

The performance of the CPU- and GPU-accelerated SDF computation for parts in Fig. \ref{figure4} is presented in Fig. \ref{figure5} (a, b) for different mesh and sample sizes. The NVIDIA Tesla K20c GPU (2,496 CUDA cores, 5GB device memory) outperforms the Intel Xeon E5-2687W CPU (32 cores, 3.10 GHz clock-rate, 64GB host memory) by average speed-up factors of $2.5-3.0\times$.\footnote{This is less than expected, because the GPU implementation suffers from extra overhead due to data transfer between host and device memories, and is suboptimal in performing conditional instructions. Much better speed-ups up to two orders of magnitude are obtained in the convolution step, as depicted next.} For this particular example, a sample size of $m = 128^3 =$ 2,097,152 and $n_{1,2} \approx 2 \times 10^5$ is adequate to capture the geometric details within the SDF field, which takes about $2$--$3$ minutes per part to precompute the SDF offline. 
The resulting SDFs of the parts in Fig. \ref{figure4} (using $\sigma := 0.5$ and $\mathfrak{p} = \lambda_2 / \lambda_1 := 3$) are plotted for their imaginary parts in Fig. \ref{figure8} (g).
The correspondence between high density regions (e.g., along the axes of cylindrical features or along the bisectors of corners of the boundary), which are analytic generalizations of virtual fixtures, is apparent.

\paragraph{Forward FFT of SDFs.}

Figure \ref{figure6} (a) presents the performance of the forward FFTs versus the sample size to map the part SDFs into the frequency domain. It is important to note that the forward FFT of part SDFs is also an offline preprocessing step, which takes negligible time compared to the previous steps (typically in the sub-millisecond range).

\subsection{Real-Time Processing} \label{sec_on}

\begin{figure}
    \centering
    \includegraphics[width=0.48\textwidth]{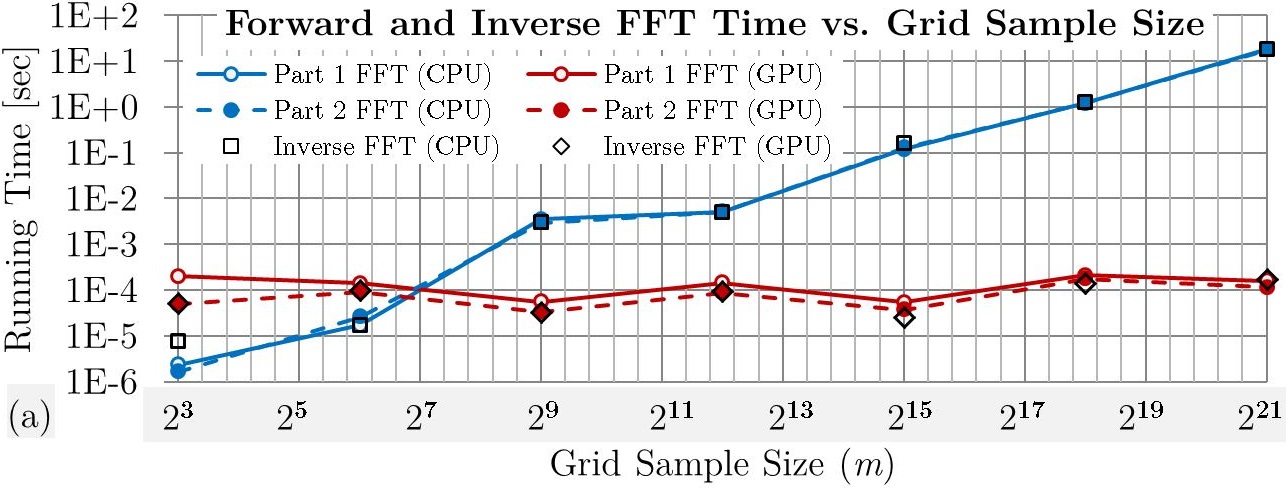}
    \includegraphics[width=0.48\textwidth]{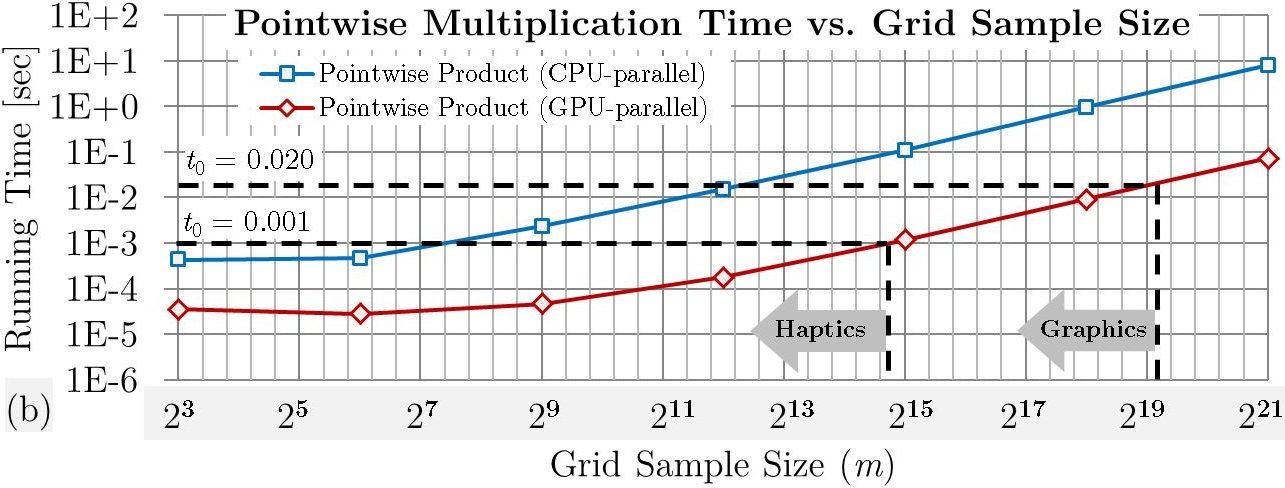}
    \caption{CPU vs. GPU performances for FFT convolution.} \label{figure6}
\end{figure}

The combination of Fourier amplitude product and inverse FFT cumulatively produces the geometric energy response for all relevant transformational configurations, which can take up to $0.1$ second for large sample sizes, as depicted in Fig. \ref{figure6} (b). However, during haptic assembly we ideally have less than $1$ millisecond, but we also need to evaluate only a single configuration at any instant of time during the virtual assembly; namely, the one corresponding to the instantaneous physical positioning of the objects. Two different strategies can be used to adapt this method to real-time applications:

\paragraph{Restricted Motion DOF.}
It is not possible in practice to precompute the energy field for a 6D grid of all possible motions. However, for most assembly scenarios the motion during the insertion phase is constrained to one or two DOF. For example, if the rotational space is limited to a finite number of permissible relative orientations, the inverse FFT for 3D translational motion can be precomputed and stored for each orientation, and queried rapidly during motion. This approach allows for computing the guidance force feedback to full accuracy, but goes against the philosophy of avoiding the multi-phase approach and manual specifications, from which we set off to pursue this method.

\paragraph{Truncated Inverse FFT.}
The major benefit of working in the frequency domain is the systematic means it provides to trade off the accuracy of physical domain representation with computation time. This provides a chance for real-time computation of translational convolution for arbitrary rotations given at any instant of assembly simulation. If $t_0$ is the amount of dedicated computation time available at each frame (e.g., ideally $t_0 \leq 0.020$ seconds for 50 Hz graphic rendering, and $t_0 \leq 0.001$ seconds for 1 kHz haptic feedback), one can always choose the maximal sample size $m_0 \ll m$ whose processing time is approximately $t_0$. Hence real-time computing is contingent upon keeping only $m' \leq m_0$ dominant modes of the SDFs in the frequency domain to approximate the convolution function (see Fig. \ref{figure3}).

Figure \ref{figure6} presents the running times of the convolution step, composed of CPU- and GPU-parallel pointwise multiplication of the mapped SDFs (plotted in panel (b)), followed by an inverse FFT to obtain the geometric energy field in the physical domain (plotted in panel (a)). Here the GPU implementation of both steps makes a crucial difference, as the inverse FFT time on the GPU stays significantly below the $1$ millisecond threshold even for sample sizes as large as $m\sim 10^6-10^7$. The pointwise multiplication is the bottleneck, whose running time determines the upperbound $m_0$ on the size of truncated Fourier expansions. For this step, CPU parallelization speeds up the multiplication process by about $25\times$ over the sequential implementation, and as depicted in Fig. \ref{figure6} the GPU implementation enhances it by an additional factor of about $100\times$ (hence an overall speed-up as much as $2,500\times$). The difference grows with the sample size, which indicates the scalability of the process.

\paragraph{Effects of Filtering.}
Figure \ref{figure8} (b--f) shows the effects of successive FFT filtering described in Section \ref{sec_filt} for different numbers of retained dominant modes $m' \leq m$ on the SDF (only imaginary part plotted on top rows). It also plots the score function on the bottom row, over a 2D section corresponding to a biaxial relative translation along the $x_1 x_2-$plane through the 6D convolution. As more frequency domain data is kept, the geometric details (e.g., pertaining to the small pairs of cylindrical pegs and holes depicted by their axis lines in panel (a)) start to emerge in the SDF shortly after $m'/m \approx 0.2\%$ in panel (d). An important observation is that the maximum score (i.e., minimum energy) configuration (denoted by B) does not change much even with very few number of frequencies in panels (b) and (c). However, the slopes and curvatures of the energy profile characterizing the forces/torques and the stiffness of combined physical and geometric constraints do change significantly. For example, for the uniaxial motion from A to B, filtering with $m'/m < 1\%$ results in a relaxed collision response and geometric guidance along BA with `soft' snapping at B, as a result of faded geometric details. By increasing the precision with $m'/m \geq 1\%$, a brisker response is imposed by larger transverse slopes along the AB trajectory with `hard' snapping at B due to sharper curvature. However, the changes are insignificant after $m'/m > 2\%$ with $\sigma = 0.5$, which enables speed-ups of two orders of magnitude by disposing of $98\%$ of the frequency data. One needs to use smaller $\sigma$ factors to capture more geometric details thus more meaningful frequencies, and larger $\mathfrak{p} = \lambda_2/\lambda_1$ to impart stronger collision response compared to collision-free geometric guidance.

\begin{figure}
    \centering
    \includegraphics[width=0.48\textwidth]{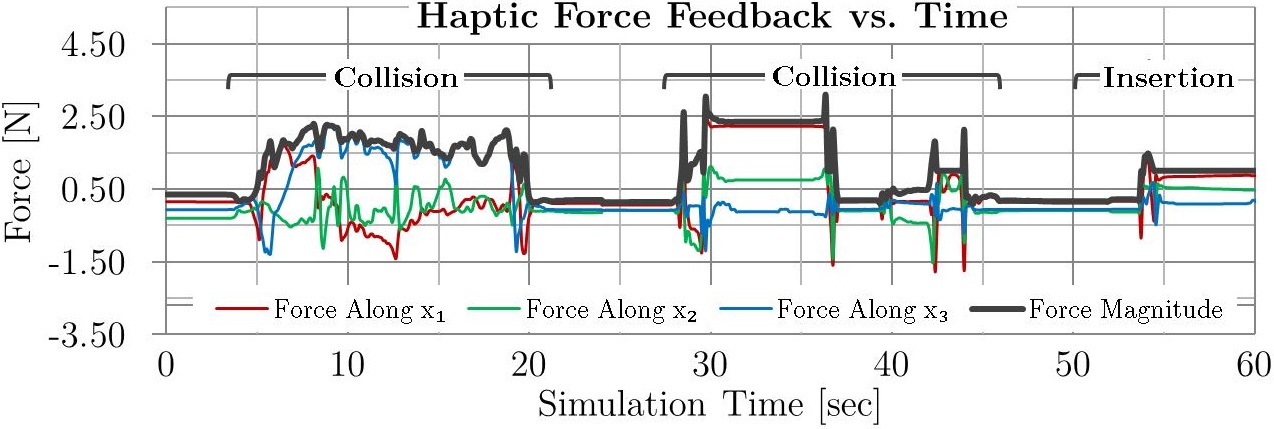}
    \includegraphics[width=0.48\textwidth]{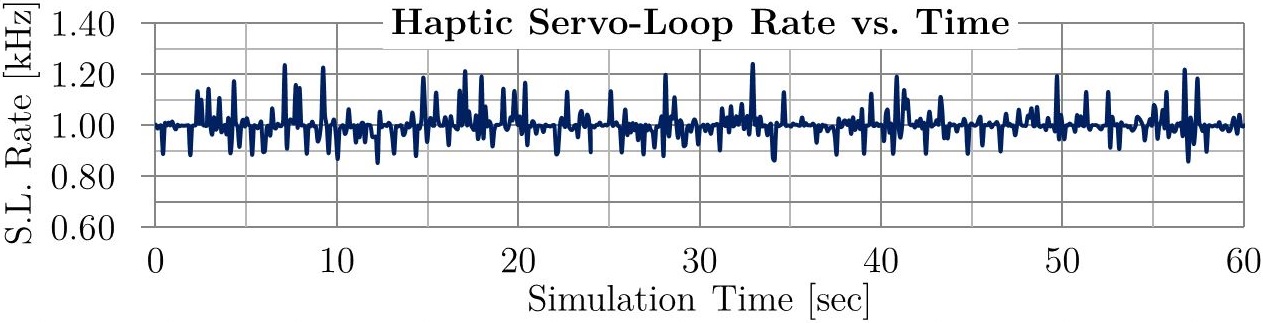}
    \caption{Performance of a haptic assembly simulation.} \label{figure7}
\end{figure}
%

\subsection{Haptic Experiment}
We test the method to assemble the pair of parts in Fig. \ref{figure4}, keeping Part 1 stationary and manipulating Part 2 using a SensAble$^\circledR$ PHANTOM$^\circledR$ Omni$^\circledR$ device (with 6 DOF input, 3 DOF output). Because the device is not capable of torque feedback, we choose to restrict the motion to the translational space, due to the observation that force response is insufficient to create a satisfactory user experience when dealing with rotational constraints. We precompute the SDFs with $\sigma = 0.5$ for fine triangular meshes of $n_1 =$ 422,272 and $n_2 =$  320,384, and a grid sample size of $m = 128^3 =$ 2,097,152.
The confinement of motion to translation only allows for exact precomputation of the convolution. However, after transferring to the Fourier domain, we keep only $m' = 16^3 =$ 4,096 (less than $0.2\%$ of $m$) dominant modes of the SDFs and zero-pad the rest.

Figure \ref{figure7} is the time plot of the haptic performance over a short assembly session, where the user explores colliding the objects from different sides and finally assembles them into the obvious configuration.
The haptic performance is consistent, always maintaining the servo-loop rate of $1.00^{+0.24}_{-0.15}$ kHz not only during free motion, but also during collision and final insertion.

\section{Conclusion}
Haptic-enabled assembly planning has been restrained for a long time from achieving its full potential, due to the seemingly contradictory requirements of handling high geometric complexity and maintaining a response rate as fast as 1 kHz.
We expanded upon our generic method in \cite{Behandish2014a,Behandish2015} to develop a unified force model for collision response and geometric guidance that applies to arbitrary geometry, by formulating the interaction energies as a convolution of the so-called SDF descriptors. Although the generality of the method inevitably imparts additional computational complexity, we demonstrated that the guidance forces and torques can be efficiently computed in the Fourier domain, where the convolution converts to pointwise product of SDF amplitudes.
We showed that the SDF shape descriptors and their Fourier expansions can be computed in a preprocessing step that takes a few minutes per part for reasonable mesh and sample sizes. For real-time computations, we showed that a very small subset of the dominant frequency domain data can be used to compute the pointwise multiplication followed by an inverse FFT. Our results confirm that such low-pass filtering of the SDF information, together with the computational power offered by the modern GPUs, enable fast evaluation of the geometric energies, forces, and torques within the available $1$ millisecond time frame during haptic assembly, with very little compromise in accuracy.
Unlike the existing approaches to collision detection or constraint management which are restricted by topological complexity (e.g., connectivity and number of holes), geometric complexity (e.g., convexity and type of surfaces), or syntactic complexity (e.g., number of triangles or voxels), our method does not impose any such restriction. Instead, it allows for a systematic trade-off between achieved fidelity and computational efficiency regardless of the input size or complexity.

The outcome of this research is a powerful paradigm that streamlines haptic assembly using spectral analysis of shape descriptors in the Fourier domain, and
opens up new promising theoretical and computational directions for VR researchers and haptics software developers.

\section{Acknowledgment}

This work was supported in part by the National Science Foundation grants CMMI-1200089, CMMI-0927105, and CNS-0927105.
The responsibility for any errors and omissions lies solely with the authors.
The identification of any commercial product or trade name does not imply endorsement or recommendation.

\bibliographystyle{asmems4}
{\scriptsize \bibliography{CDL-TR-16-01}}

\begin{thebibliography}{10}

\bibitem{Behandish2015a}
Behandish, M., and Ilie\c{s}, H.~T., 2015.
\newblock ``Haptic assembly using skeletal densities and {F}ourier
  transforms''.
\newblock In Proceedings of the 2015 ASME International Design Engineering
  Technical Conferences and Computers and Information in Engineering Conference
  (IDETC/CIE'2015).

\bibitem{Bullinger1999}
Bullinger, H.~J., Breining, R., and Bauer, W., 1999.
\newblock ``Virtual prototyping--state of the art in product design''.
\newblock In 26th International Conference on Computers and Industrial
  Engineering, pp.~103--107.

\bibitem{Wang2002}
Wang, G.~G., 2002.
\newblock ``Definition and review of virtual prototyping''.
\newblock {\em Journal of Computing and Information Science in Engineering,
  {\bf 2}}(3), pp.~232--236.

\bibitem{Deviprasad2003}
Deviprasad, T., and Kesavadas, T., 2003.
\newblock ``Virtual prototyping of assembly components using process
  modeling''.
\newblock {\em Journal of Manufacturing Systems, {\bf 22}}(1), pp.~16--27.

\bibitem{Bordegoni2006}
Bordegoni, M., Colombo, G., and Formentini, L., 2006.
\newblock ``Haptic technologies for the conceptual and validation phases of
  product design''.
\newblock {\em Computers and Graphics, {\bf 30}}(3), pp.~377--390.

\bibitem{Seth2006}
Seth, A., Su, H.~J., and Vance, J.~M., 2006.
\newblock ``{SHARP}: A system for haptic assembly and realistic prototyping''.
\newblock In Proceedings of the 2006 ASME International Design Engineering
  Technical Conferences and Computers and Information in Engineering Conference
  (IDETC/CIE'2006), pp.~905--912.

\bibitem{Seth2008}
Seth, A., Su, H.~J., and Vance, J.~M., 2008.
\newblock ``Development of a dual-handed haptic assembly system: {SHARP}''.
\newblock {\em Journal of Computing and Information Science in Engineering,
  {\bf 8}}(4), pp.~1--8.

\bibitem{Gomes1999}
Gomes~de Sa, A., and Zachmann, G., 1999.
\newblock ``Virtual reality as a tool for verification of assembly and
  maintenance processes''.
\newblock {\em Computers and Graphics, {\bf 23}}(3), pp.~389--403.

\bibitem{Volkov2001}
Volkov, S., and Vance, J.~M., 2001.
\newblock ``Effectiveness of haptic sensation for the evaluation of virtual
  prototypes''.
\newblock {\em Journal of Computing and Information Science in Engineering,
  {\bf 1}}(2), pp.~123--128.

\bibitem{Seth2011}
Seth, A., Vance, J.~M., and Oliver, J.~H., 2011.
\newblock ``Virtual reality for assembly methods prototyping: A review''.
\newblock {\em Virtual Reality, {\bf 15}}(1), pp.~5--20.

\bibitem{Vance2011}
Vance, J.~M., and Dumont, G., 2011.
\newblock ``A conceptual framework to support natural interaction for virtual
  assembly tasks''.
\newblock In Proceedings of the 2011 ASME World Conference on Innovative
  Virtual Reality, pp.~273--278.

\bibitem{Perret2013}
Perret, J., Kneschke, C., Vance, J.~M., and Dumont, G., 2013.
\newblock ``Interactive assembly simulation with haptic feedback''.
\newblock {\em Assembly Automation, {\bf 33}}(3), pp.~214--220.

\bibitem{Behandish2014a}
Behandish, M., and Ilie\c{s}, H.~T., 2014.
\newblock ``Peg-in-hole revisited: A generic force model for haptic assembly''.
\newblock In Proceedings of the 2014 ASME International Design Engineering
  Technical Conferences and Computers and Information in Engineering Conference
  (IDETC/CIE'2014).

\bibitem{Behandish2015}
Behandish, M., and Ilie\c{s}, H.~T., 2015.
\newblock ``Peg-in-hole revisited: A generic force model for haptic assembly''.
\newblock {\em Journal of Computing and Information Science in Engineering,
  {\bf $~$}}(To Appear).

\bibitem{Lysenko2010}
Lysenko, M., Nelaturi, S., and Shapiro, V., 2010.
\newblock ``Group morphology with convolution algebras''.
\newblock In Proceedings of the 2010 ACM Symposium on Solid and Physical
  Modeling (SPM'2010), pp.~11--22.

\bibitem{Lysenko2013}
Lysenko, M., 2013.
\newblock ``{F}ourier collision detection''.
\newblock {\em International Journal of Robotics Research, {\bf 32}}(4),
  pp.~483--503.

\bibitem{Behandish2015d}
Behandish, M., and Ilie\c{s}, H.~T., 2015.
\newblock ``Analytic methods for geometric modeling via spherical
  decomposition''.
\newblock {\em Computer-Aided Design, Special Issue on the 2015 SIAM Conference
  on Geometric and Physical Modeling (GD/SPM'2015)}.

\bibitem{Roerdink2000}
Roerdink, J. B. T.~M., 2000.
\newblock ``Group morphology''.
\newblock {\em Pattern Recognition, {\bf 33}}(6), pp.~877--895.

\bibitem{Nelaturi2011}
Nelaturi, S., and Shapiro, V., 2011.
\newblock ``Configuration products and quotients in geometric modeling''.
\newblock {\em Computer-Aided Design, {\bf 43}}(7), pp.~781--794.

\bibitem{Behandish2014}
Behandish, M., and Ilie\c{s}, H.~T., 2014.
\newblock Shape complementarity analysis for objects of arbitrary shape.
\newblock Tech. rep., University of Connecticut.

\bibitem{Lozano-Perez1983}
Lozano-Perez, T., 1983.
\newblock ``Spatial planning: A configuration space approach''.
\newblock {\em IEEE Transactions on Computers, {\bf C-32}}(2), pp.~108--120.

\bibitem{Kavraki1995}
Kavraki, L.~E., 1995.
\newblock ``Computation of configuration-space obstacles using the fast
  {F}ourier transform''.
\newblock {\em IEEE Transactions on Robotics and Automation, {\bf 11}}(3),
  pp.~408--413.

\bibitem{Chan2009}
Chan, L. S.~H., and Choi, K.~S., 2009.
\newblock ``Integrating {PhysX} and {OpenHaptics}: Efficient force feedback
  generation using physics engine and haptic devices''.
\newblock In Proceedings of the 2009 Joint Conferences on Pervasive Computing
  (JCPC'2009),, pp.~853--858.

\bibitem{Sagardia2014}
Sagardia, M., Stouraitis, T., and Silva, J. L.~e., 2014.
\newblock ``A new fast and robust collision detection and force computation
  algorithm applied to the physics engine bullet: Method, integration, and
  evaluation''.
\newblock In Proceedings of the 2014 Conference and Exhibition of the European
  Association of Virtual and Augmented Reality (EuroVR'2014).

\bibitem{Renouf2005}
Renouf, M., Acary, V., and Dumont, G., 2005.
\newblock ``{3D} frictional contact and impact multibody dynamics. a comparison
  of algorithms suitable for real-time applications''.
\newblock In ECCOMAS Thematic Conference on Mutlibody Dynamics.

\bibitem{Tching2008}
Tching, L., and Dumont, G., 2008.
\newblock ``Haptic simulations based on non-smooth dynamics for rigid-bodies''.
\newblock In Proceedings of the 15th ACM Symposium on Virtual Reality Software
  and Technology (VRST'2008), pp.~87--90.

\bibitem{Lin1998}
Lin, M., and Gottschalk, S., 1998.
\newblock ``Collision detection between geometric models: A survey''.
\newblock In Proceedings of the 1998 IMA Conference on Mathematics of Surfaces,
  Vol.~1, pp.~37--56.

\bibitem{Jimenez2001}
Jimenez, P., Thomas, F., and Torras, 2001.
\newblock ``{3D} collision detection: A survey''.
\newblock {\em Computers and Graphics, {\bf 25}}(2), pp.~269--285.

\bibitem{Kockara2007}
Kockara, S., Halic, T., Iqbal, K., Bayrak, C., and Rowe, R., 2007.
\newblock ``Collision detection: A survey''.
\newblock In Proceedings of the 2007 IEEE International Conference on Systems,
  Man and Cybernetics (ISIC'2007), pp.~4046--4051.

\bibitem{Hasegawa2003}
Hasegawa, S., and Fujii, N., 2003.
\newblock ``Real-time rigid body simulation based on volumetric penalty
  method''.
\newblock In Proceedings of the 11th Symposium on Haptic Interfaces for Virtual
  Environment and Teleoperator Systems, pp.~326--332.

\bibitem{Hasegawa2004}
Hasegawa, S., and Sato, M., 2004.
\newblock ``Real-time rigid body simulation for haptic interactions based on
  contact volume of polygonal objects''.
\newblock {\em Computer Graphics Forum, {\bf 23}}(3), pp.~529--538.

\bibitem{Lotstedt1984}
L{\"o}tstedt, P., 1984.
\newblock ``Numerical simulation of time-dependent contact and friction
  problems in rigid body mechanics''.
\newblock {\em SIAM Journal on Scientific and Statistical Computing, {\bf
  5}}(2), pp.~370--393.

\bibitem{Stewart2000}
Stewart, D.~E., 2000.
\newblock ``Rigid-body dynamics with friction and impact''.
\newblock {\em SIAM Review, {\bf 42}}(1), pp.~3--39.

\bibitem{Marcelino2003}
Marcelino, L., Murray, N., and Fernando, T., 2003.
\newblock ``A constraint manager to support virtual maintainability''.
\newblock {\em Computers and Graphics, {\bf 27}}(1), pp.~19--26.

\bibitem{Murray2004}
Murray, N., and Fernando, T., 2004.
\newblock ``An immersive assembly and maintenance simulation environment''.
\newblock In Proceedings of the 2004 IEEE International Symposium on
  Distributed Simulation and Real-Time Applications (DS-RT'2004), pp.~159--166.

\bibitem{Iacob2008}
Iacob, R., Mitrouchev, P., and Leon, J.~C., 2008.
\newblock ``Contact identification for assembly-disassembly simulation with a
  haptic device''.
\newblock {\em The Visual Computer, {\bf 24}}(11), pp.~973--979.

\bibitem{Iacob2011}
Iacob, R., Mitrouchev, P., and Leon, J.~C., 2011.
\newblock ``Assembly simulation incorporating component mobility modelling
  based on functional surfaces''.
\newblock {\em International Journal on Interactive Design and Manufacturing,
  {\bf 5}}(2), pp.~119--132.

\bibitem{Boussuge2012}
Boussuge, F., L{\'e}on, J.~C., Hahmann, S., and Fine, L., 2012.
\newblock ``An analysis of {DMU} transformation requirements for structural
  assembly simulations''.
\newblock In Proceedings of the 8th International Conference on Engineering
  Computational Technology.

\bibitem{Mirtich1998}
Mirtich, B., 1998.
\newblock ``{V}-{C}lip: Fast and robust polyhedral collision detection''.
\newblock {\em ACM Transactions on Graphics (TOG), {\bf 17}}(3), pp.~177--208.

\bibitem{Ehmann2000}
Ehmann, S.~A., and Lin, M.~C., 2000.
\newblock ``Accelerated proximity queries between convex polyhedra by
  multi-level {V}oronoi marching''.
\newblock In Proceedings of the 2000 IEEE/RSJ International Conference on
  Intelligent Robots and Systems (IROS'2000), Vol.~3.

\bibitem{Ehmann2001}
Ehmann, S.~A., and Lin, M.~C., 2001.
\newblock ``Accurate and fast proximity queries between polyhedra using convex
  surface decomposition''.
\newblock {\em Computer Graphics Forum, {\bf 20}}(3), pp.~500--511.

\bibitem{Gregory1999}
Gregory, A., Lin, M.~C., Gottschalk, S., and Taylor, R., 1999.
\newblock ``{H-COLLIDE}: A framework for fast and accurate collision detection
  for haptic interaction''.
\newblock In Proceedings of the 1999 IEEE Virtual Reality Conference (VR'1999),
  pp.~38--45.

\bibitem{Gregory2005}
Gregory, A., Lin, M.~C., Gottschalk, S., and Taylor, R., 2005.
\newblock ``Fast and accurate collision detection for haptic interaction using
  a three degree-of-freedom force-feedback device''.
\newblock In Proceedings of the ACM SIGGRAPH'2005 Courses.

\bibitem{Gottschalk1996}
Gottschalk, S., Lin, M.~C., and Manocha, D., 1996.
\newblock ``{OBBT}ree: A hierarchical structure for rapid interference
  detection''.
\newblock In Proceedings of the 23rd Annual Conference on Computer Graphics and
  Interactive Techniques, pp.~171--180.

\bibitem{McNeely2005}
McNeely, W.~A., Puterbaugh, K.~D., and Troy, J.~J., 2005.
\newblock ``Six degree-of-freedom haptic rendering using voxel sampling''.
\newblock In Proceedings of the ACM SIGGRAPH'2005 Courses.

\bibitem{McNeely2006}
McNeely, W.~A., Puterbaugh, K.~D., and Troy, J.~J., 2006.
\newblock ``Voxel-based 6-{DOF} haptic rendering improvements''.
\newblock {\em Haptics-e, {\bf 3}}(7).

\bibitem{Barbic2007}
Barbi\v{c}, J., and James, D., 2007.
\newblock ``Time-critical distributed contact for 6-{DOF} haptic rendering of
  adaptively sampled reduced deformable models''.
\newblock In Proceedings of the ACM SIGGRAPH'2007/Eurographics Symposium on
  Computer Animation, pp.~171--180.

\bibitem{Sagardia2008}
Sagardia, M., Hulin, T., Preusche, C., and Hirzinger, 2008.
\newblock ``Improvements of the {V}oxmap-{P}oint{S}hell algorithm--fast
  generation of haptic data-structures''.
\newblock In 53rd Internationales Wissenschaftliches Kolloquium Technische
  Universität Ilmenau.

\bibitem{Hubbard1996}
Hubbard, P.~M., 1996.
\newblock ``Approximating polyhedra with spheres for time-critical collision
  detection''.
\newblock {\em ACM Transac, {\bf 15}}(3), pp.~179--210.

\bibitem{Bradshaw2004}
Bradshaw, G., and O'Sullivan, C., 2004.
\newblock ``Adaptive medial-axis approximation for sphere-tree construction''.
\newblock {\em ACM Transactions on Graphics (TOG), {\bf 23}}(1), pp.~1--26.

\bibitem{Weller2009}
Weller, R., and Zachmann, G., 2009.
\newblock ``Inner sphere trees for proximity and penetration queries''.
\newblock In Proceedings of the 2009 Robotics: Science and Systems Conference
  (RSS'2009), Vol.~2.

\bibitem{Weller2011}
Weller, R., and Zachmann, G., 2011.
\newblock ``Inner sphere trees and their application to collision detection''.
\newblock In {\em Virtual Realities}, G.~Brunnett, S.~Coquillart, and G.~Welch,
  eds. Springer Vienna, pp.~181--201.

\bibitem{Ruffaldi2008}
Ruffaldi, E., Morris, D., Barbagli, F., Salisbury, K., and Bergamasco, M.,
  2008.
\newblock ``Voxel-based haptic rendering using implicit sphere trees''.
\newblock In Proceedings of the 16th International Symposium on Haptic
  Interfaces for Virtual Environment and Teleoperator Systems, pp.~319--325.

\bibitem{Weller2009a}
Weller, R., and Zachmann, G., 2009.
\newblock ``A unified approach for physically-based simulations and haptic
  rendering''.
\newblock In Proceedings of the ACM SIGGRAPH'2009 Symposium on Video Games,
  pp.~151--159.

\bibitem{Weller2009b}
Weller, R., and Zachmann, G., 2009.
\newblock ``Stable 6-{DOF} haptic rendering with inner sphere trees''.
\newblock In Proceedings of the 2009 ASME International Design Engineering
  Technical Conferences and Computers and Information in Engineering Conference
  (IDETC/CIE'2009).

\bibitem{Seth2007}
Seth, A., Vance, J.~M., and Oliver, J.~H., 2007.
\newblock ``Combining geometric constraints with physics modeling for virtual
  assembly using {SHARP}''.
\newblock In Proceedings of the 2007 ASME International Design Engineering
  Technical Conferences and Computers and Information in Engineering Conference
  (IDETC/CIE'2007).

\bibitem{Seth2010}
Seth, A., Vance, J.~M., and Oliver, J.~H., 2010.
\newblock ``Combining dynamic modeling with geometric constraint management to
  support low clearance virtual manual assembly''.
\newblock {\em Journal of Mechanical Design, {\bf 132}}(8), pp.~081002--1--7.

\bibitem{Jayaram1999}
Jayaram, S., Jayaram, U., Wang, Y.and~Tirumali, H., Lyons, K., and Hart, P.,
  1999.
\newblock ``{VADE}: A virtual assembly design environment''.
\newblock {\em Journal of Computer Graphics and Applications, {\bf 19}}(6),
  pp.~44--50.

\bibitem{Wan2004}
Wan, H., Gao, S., Peng, Q., Dai, G., and Zhang, F., 2004.
\newblock ``{MIVAS}: A multi-modal immersive virtual assembly system''.
\newblock In Proceedings of the 2004 ASME International Design Engineering
  Technical Conferences and Computers and Information in Engineering Conference
  (IDETC/CIE'2004), pp.~113--122.

\bibitem{Tching2010a}
Tching, L., Dumont, G., and Perret, J., 2010.
\newblock ``Interactive simulation of {CAD} models assemblies using virtual
  constraint guidance''.
\newblock {\em International Journal on Interactive Design and Manufacturing,
  {\bf 4}}(2), pp.~95--102.

\bibitem{Rosenberg1993}
Rosenberg, L.~B., 1993.
\newblock ``Virtual fixtures: Perceptual tools for telerobotic manipulation''.
\newblock In Proceedings of the 1993 IEEE Virtual Reality Annual International
  Symposium, pp.~76--82.

\bibitem{Katznelson2004}
Katznelson, Y., 2004.
\newblock {\em An Introduction to Harmonic Analysis}, 3rd edition~ed.
\newblock Cambridge University Press.

\bibitem{Cooley1965}
Cooley, J.~W., T.~J., 1965.
\newblock ``An algorithm for the machine calculation of complex {F}ourier
  series''.
\newblock {\em Mathematics of CSomputation, {\bf 19}}(90), pp.~297--301.

\bibitem{Requicha1977a}
Requicha, A.~G., 1980.
\newblock Mathematical models of rigid solid objects.
\newblock Production Automation Project, Technical Memo. No. 28, University of
  Rochester.

\bibitem{Chazal2004}
Chazal, F., and Soufflet, R., 2004.
\newblock ``Stability and finiteness properties of medial axis and skeleton''.
\newblock {\em Journal of Dynamical and Control Systems, {\bf 10}}(2),
  pp.~149--170.

\bibitem{Klein2009}
Klein, F., 2009.
\newblock ``A new approach to point membership classification in {B}-rep
  solids''.
\newblock In {\em Mathematics of Surfaces XIII}. Springer Berlin Heidelberg,
  pp.~235--250.

\bibitem{Requicha1980a}
Requicha, A.~G., 1980.
\newblock Representations of rigid solid objects.
\newblock Production Automation Project, Technical Memo. No. 29, University of
  Rochester.

\bibitem{Schoberl1997}
Sch{\"o}berl, J., 1997.
\newblock ``{NETGEN}: An advancing front {2D}/{3D}-mesh generator based on
  abstract rules''.
\newblock {\em Computing and Visualization in Science, {\bf 1}}(1), pp.~41--52.

\bibitem{Hoff1999}
Hoff~III, K.~E., Culver, T., Keyser, J., Lin, M., and Manocha, D., 1999.
\newblock ``Fast computation of generalized {V}oronoi diagrams using graphics
  hardware''.
\newblock In Proceedings of the 26th Annual Conference on Computer Graphics and
  Interactive Techniques, pp.~277--286.

\bibitem{Schling2011}
Schling, B., 2011.
\newblock {\em The {B}oost {C}++ Libraries}.
\newblock XML Press.

\bibitem{Frigo2005}
Frigo, M., and Johnson, S.~G., 2005.
\newblock ``The design and implementation of {FFTW3}''.
\newblock {\em Proceedings of the IEEE, Special issue on Program Generation,
  Optimization, and Platform Adaptation, {\bf 93}}(2), pp.~216--231.

\end{thebibliography}

\end{document}